\documentstyle[epsfig,graphics]{mn}

\begin{document}

\newcommand{\etal}{{\em et al.}\ }                   
\newcommand{\ie}{{\em i.e. }}
\newcommand{\eg}{{\em e.g. }}
\newcommand{\micron}{\mbox{\,${\mu}$m} }             
\newcommand{\Msolar}{\mbox{\,$\rm M_{\odot}$}}        
\newcommand{\Lsolar}{\mbox{\,$\rm L_{\odot}$}}        
\newcommand{\asec}{^{\prime\prime}}
\newcommand{\h}{{\sc hst\,}}
\newcommand{\irc}{{\sc ircam3\,}}
\newcommand{\wf}{{\sc wfpc2\,}}
\newcommand{\psf}{{\sc psf\,}}
\hyphenation{infra-red}    
\hyphenation{inter-stellar}
\renewcommand{\baselinestretch}{1.5}
\newcommand{\spsc}[2]{\mbox{$\rm #1^{#2}$}}     
\newcommand{\sbsc}[2]{\mbox{$\rm #1_{#2}$}}     
\newcommand{\ang}{\mbox{$\rm \AA$}}
\newcommand{\unit}[1]{\; \rm #1}

\newcommand{\xs}{$\chi^{2}$}


\title[The ages of quasar host galaxies.]{The ages of quasar host galaxies.}
\author[L.A. Nolan \etal.]
{L.A. Nolan$^{1}$, J.S. Dunlop$^{1}$, M.J. Kukula$^{1}$, D. H.
Hughes$^{1,2}$, T. Boroson$^{3}$ $\&$  R. Jimenez$^{1}$. 
\\
$^{1}$Institute for Astronomy, University of Edinburgh, Blackford Hill,
Edinburgh, EH9~3HJ\\
$^{2}$INOAE, Apartado Postal 51 y 216, 72000, Puebla, Pue., Mexico.\\
$^{3}$NOAO, PO Box 26732, Tucson, Arizona 85726$-$6732, U.S.A.}

\date{Submitted for publication in MNRAS}

\maketitle
  
\begin{abstract}

We present the results of fitting deep off-nuclear optical spectra of
radio-quiet quasars, radio-loud quasars and radio galaxies at $z \simeq
0.2$ with evolutionary synthesis models of galaxy evolution. Our aim was
to determine the age of the dynamically dominant stellar populations in the host
galaxies of these three classes of powerful AGN. Some of our spectra
display residual nuclear contamination at the shortest wavelengths, but
the detailed quality of the fits longward of the 4000\AA\ break provide
unequivocal proof, if further proof were needed, that quasars lie in
massive galaxies with (at least at $z \simeq 0.2$) evolved stellar
populations. By fitting a two-component model we have separated the very
blue (starburst and/or AGN contamination) from the redder underlying
spectral energy distribution, and find that the hosts of all three
classes of AGN are dominated by old stars of age $8 - 14$ Gyr. If the blue
component is attributed to young stars, we find that, at most, 1\% of the
visible baryonic mass of these galaxies is involved in star-formation activity at the epoch
of observation, at least over the region sampled by our spectroscopic
observations. 
These results strongly support the conclusion reached
by McLure et al. (1999) that the
host galaxies of luminous quasars are massive ellipticals which have formed
by the epoch of peak quasar activity at $z \simeq 2.5$.

\end{abstract}


\section{Introduction}

Determining the nature of the host galaxies of powerful active galactic 
nuclei (AGN) is of importance not only for improving our understanding 
of different manifestations of AGN activity, but also for 
exploring possible relationships between nuclear activity and the 
evolution of massive galaxies. The recent affirmation that black-hole mass
appears approximately proportional to spheroid mass in nearby inactive glaxies
(Maggorian et al. 1998) has further strengthened the motivation for 
exploring the link between active AGN and the dynamical and spectral 
properties of their hosts. 

Over the last few years, improvements in imaging resolution offered by
both space-based and ground-based optical/infrared telescopes has stimulated 
a great deal of research activity aimed at determining the basic structural 
parameters ({\it i.e.} luminosity, size and morphological type) of the hosts 
of radio-loud quasars, radio-quiet quasars, and lower-luminosity X-ray-selected
and optically-selected AGN ({\it e.g.} Disney et al. 1995; 
Hutchings \& Morris 1995; Bahcall et al. 1997; Hooper et al. 1997; 
McLure et al. 1999, Schade et al. 2000). However, relatively little corresponding effort has been invested in spectroscopic investigations of AGN 
hosts, despite the fact that this offers an independent way of classifying 
these galaxies, as well as a means of estimating the age of their 
stellar populations.

Our own work in this field to date has focussed on the investigation of
the hosts of matched samples of radio-quiet quasars (RQQs), radio-loud quasars
(RLQs) and radio galaxies (RGs) at relatively modest redshift ($z = 0.2$).
Details of these samples can be found in Dunlop et al. (1993). In
brief, the sub-samples of quasars (i.e. RQQs and RLQs) have been selected
to be indistinguishable in terms of their two-dimensional distribution
on the $V-z$ plane, while the sub-samples of radio-loud AGN (i.e. RLQs
and RGs) have been selected to be indistinguishable in terms of their
two-dimensional distribution on the $P_{5GHz} - z$ plane (as well as
having indistinguishable spectral-index distributions).
Deep infrared imaging of these samples (Dunlop et al. 1993; Taylor et al. 1996)
has recently been complemented by deep WFPC2 HST optical imaging (McLure
et al. 1999), the final results of which are reported by Dunlop et al. 
(2000). As well as demonstrating that, dynamically, the hosts of all three
types of luminous AGN appear indistinguishable from normal ellipticals, this 
work has enabled us to deduce crude spectral information on the host galaxies
in the form of optical-infrared colours. However, broad-baseline colour
information can clearly be most powerfully exploited if combined with 
detailed optical spectroscopy.
Over the past few years we have therefore attempted to complement 
our imaging studies with a programme of 
deep optical off-nuclear spectroscopy of this same sample of AGN.

Details of the observed samples, spectroscopic observations, and the 
basic properties of the observed off-nuclear spectra are given in a companion
paper (Hughes et al. 2000). As discussed by Hughes et al. (2000), the key 
feature of this study (other than its size, depth, and sample control) is 
that we have endeavoured to obtain spectra from positions further off-nucleus 
($\simeq 5$ arcsec) than previous workers, in an attempt to better 
minimize the need for accurate subtraction of contaminating nuclear light. 
This approach was made possible by our deep infrared imaging data, which 
allowed us to select slit positions $\simeq 5$ arcsec off-nucleus 
which still intercepted the brighter isophotes of the host galaxies (slit
positions are shown, superimposed on the infrared images, in Hughes et al.
2000).

In this paper we present the results of attempting to fit the
resulting  off-nuclear
spectra with evolutionary synthesis models of galaxy evolution. Our 
primary aim was to determine whether, in each host galaxy, the optical
spectrum could be explained by the same model as the optical-infrared
colour, and, if so, to derive an estimate of the age of the dynamically 
dominant stellar population. However, this study also offered the prospect
of determining whether the hosts of different classes of AGN differ 
in terms of their more recent star-formation activity.   

It is worth noting that our off-nuclear spectroscopy obviously does not
enable us to say anything about the level of star-formation activity in
the nucleus of a given host galaxy. Rather, any derived estimates of the
level of star-formation activity refer to the region probed by our
observations $\simeq 5$ arcsec off-nucleus. However, given the large
scalelengths of the hosts, their relatively modest redshift, and the fact
that our spectra are derived from long-slit observations, it is
reasonable to regard our conclusions as applying to fairly typical
regions, still located well within the bulk of the host galaxies under
investigation.
 
The layout of this paper is as follows. In section 2 we provide details of 
the adopted models, and how they have been fitted to the data. The results are
presented in section 3, along with detailed notes on the fitting of 
individual spectra. The implications of the model fits are then discussed 
in section 4, focussing on a comparison of the typical derived host-galaxy 
ages in the three AGN subsamples. Finally our conclusions are summarized 
in section 5. 
The detailed model fits, along with corresponding chi-squared plots are 
presented in Appendix A.

\section{Spectral fitting.}

The stellar population synthesis models adopted for age-dating the AGN host 
galaxy stellar populations are the solar metallicity, instantaneous starburst 
models of Jimenez et al. (2000). We have endeavoured to fit each off-nuclear
host galaxy optical spectrum by a combination of two single starburst 
components. For the first component, age is fitted as a free
parameter, while for the second component the age is fixed at 0.1 Gyr,
and the normalization relative to the first component is the only free
parameter. This dual-component approach was adopted because single-age
models are not able to adequately represent the data, and because it allows
the age of the dynamically dominant component to be determined in a way which
is not overly reliant on the level of ultraviolet flux which might be
contributed either by a recent burst of star-formation, or by
contamination of the slit by scattered light from the quasar nucleus.
After experimentation it was found that the spectral shape of the 
blue light was better represented by the 0.1 Gyr-old (solar metallicity) model 
of Jimenez than by models of greater (intermediate) age (e.g. 1 Gyr). 
Moreover, the further addition of a third intermediate age component, 
$\simeq$ 1 Gyr, did not significantly improve the quality of the fits
achieved.

In general, our data are of insufficient quality to tell whether the
component which dominates at $\lambda \simeq 3000$\AA\ really is due to
young stars, or is produced by direct or scattered quasar light. However, 
while the origin of the blue light is obviously of some interest, it has little
impact on the main results presented in this paper, which refer to the
age of the dynamically dominant stellar population which dominates the
spectrum from longward of the 4000\AA\ break through to the near infrared. The robustness of the age determination of the dominant population is demonstrated by comparing the results of the two stellar-population model fit with the results of a fit allowi
ng for a nuclear contribution as well as the two stellar populations.

The model parameters determined were therefore 
the age of the dominant stellar population, 
which we can reasonably call the age of the galaxy, and the fraction, by 
(visible) baryonic mass, of the 0.1 Gyr component. For the fit including a nuclear contribution, the fraction of the total flux contributed by quasar light was a third parameter. The red end of each SED was further 
constrained by fitting $R-K$ colour simultaneously with the optical
spectral energy distribution. The fitting process is described below.

First, the observed off-nuclear spectra were corrected for redshift and transformed to the rest frame. They were then rebinned to the 
spectral resolution of the model spectra. The rebinned flux is then the mean 
flux per unit wavelength in the new bin and the statistical error on each 
new bin is the standard error in this mean. The data were normalised to a 
mean flux per unit wavelength of unity across the wavelength range 5020 $-$ 
5500 \ang.

The two-component model was built from the instantaneous-burst stellar-population synthesis model SEDs, so that
  $$	F_{\lambda,age,\alpha} = const (\alpha f_{\lambda,0.1} + (1-\alpha)f_{\lambda,age})  $$
where $f_{\lambda,age}$ is the mean flux per unit wavelength in the bin 
centred on wavelength $\lambda$ for a single burst model of $age$ 
Gyr, $\alpha$ is the fraction by mass of the young (0.1 Gyr) component, 
and $F_{\lambda,age,\alpha}$ is then the new, mean twin stellar-population 
flux per unit wavelength in the bin centred on wavelength $\lambda$ for a 
model of $age$ Gyr. 
This composite spectrum was then normalised in the same way as the observed 
spectra.

A \xs\ fit was used to determine the age of the older stellar population and 
the mass-fraction, $\alpha$, of the younger population, for each host galaxy 
in the sample. The whole parameter space was searched, with the best-fit 
values quoted being those parameter values at the point on the grid with the 
minimum calculated  \xs . The normalization of the model spectra was
allowed to float during the fitting 
process, to allow the best-fitting continuum shape to be determined in an
unbiased way.

The models were fitted across the observed (rest-frame) spectral range, within the wavelength ranges listed in Table 1. 
As a result of the optimisation of the instruments with which the objects 
were observed, some of the galaxy spectra contain a `splice' region where 
the red and blue end of the spectrum have been observed separately and then 
joined together (see Hughes \etal (2000) for details). These splice 
regions, defined in Table 1, were masked out of the fit in order to guard
against the fitting procedure being dominated by data-points whose flux
calibration was potentially less robust. The main emission 
lines, present due to nuclear light contamination or nebular emission
from within the host, were also masked out, over the wavelength ranges 
given in Table 2.

The fit including a nuclear component was carried out in the same way, with the model flux in this case being
  $$	F_{\lambda,age,\alpha,\eta} = const (\alpha f_{\lambda,0.1} + (1-\alpha)f_{\lambda,age} + \eta f0054_{\lambda})  $$
where $\eta$ is the fraction contributed to the total model flux by the nucleus, and 
$f0054_{\lambda}$ is the observed flux of the nucleus of the radio quiet quasar 0054+144. $\alpha$ is the fraction by mass of the total stellar population contributed by the 0.1 Gyr population. The wavelength range of the observed nuclear flux of 0054+144
 is 3890$-$6950 \ang. This was extended to  3500$-$8500 \ang\ by smooth extrapolation over the wavelength ranges 3500$-$3890 \ang\ and 6950$-$8500 \ang\ in order to carry out the \xs\ fit across the full wavelength range of the observed off-nuclear spectr
a.

\begin{table*}

\caption{Wavelength ranges of observed spectra, and splice regions which
were excluded from the fitting process. All wavelengths are in the observed 
frame. M4M denotes the Mayall 4m Telescope at Kitt Peak, and WHT denotes 
the 4.2m William Herschel Telescope on La Palma (see Hughes \etal (2000)
for observational details).}

\begin{tabular}{lccc}

\hline

   IAU name & Telescope & Wavelength range / \ang & Masked splice region / \ang       \\

\hline

\multicolumn{4}{c}{ \large \it Radio Loud Quasars}	 \\ 
  	
\hline

 0137+012 & M4M & 3890$-$6950 &	 	\\   
 0736+017 & M4M & 3890$-$6950 &		\\  
	  & WHT & 3500$-$8000 & 6050$-$6150 \\
 1004+130 & WHT & 3500$-$8000 & 6050$-$6150	\\  
 1020$-$103 & M4M & 3890$-$6950 &		\\  
 1217+023 & WHT & 3500$-$7500 & 6000$-$6100	\\  
 2135$-$147 & WHT & 3500$-$8500 & 6000$-$6100	\\  
 2141+175 & WHT & 3500$-$8500 & 6000$-$6100	\\  
 2247+140 & M4M & 3890$-$6950 &		\\  
	  & WHT & 3500$-$7500 & 6050$-$6150	\\
 2349$-$014 & WHT & 3500$-$8300 & 6000$-$6100 \\	

\hline

\multicolumn{4}{c}{ \large \it Radio Quiet Quasars}  	\\

\hline

 0054+144 & M4M & 3890$-$6950 &		\\
	  & WHT & 3500$-$8000 & 6000$-$6100	\\
 0157+001 & M4M & 3890$-$6950 &		\\
 	  & WHT & 3500$-$8000 & 6050$-$6150 \\
 0204+292 & WHT & 3500$-$8000 & 6050$-$6150 \\
 0244+194 & WHT & 3500$-$8500 & 6000$-$6100	\\  
 0923+201 & WHT & 3500$-$7000 &		\\
 1549+203 & WHT & 3500$-$8000 & 6050$-$6150	\\  
 1635+119 & WHT & 3500$-$7800 & 6000$-$6100	\\  
 2215$-$037 & WHT & 3500$-$8500 & 6000$-$6100	\\ 
 2344+184 & M4M & 3890$-$6950 &		\\
	  & WHT & 3500$-$8500 & 6000$-$6100	\\

\hline

\multicolumn{4}{c}{\large \it Radio Galaxies}                        \\

\hline

 0230$-$027 & WHT & 3500$-$8500 & 6000$-$6100	\\ 
 0345+337 & WHT & 3500$-$8500 & 6000$-$6100	\\  
 0917+459 & WHT & 3500$-$8500 & 6050$-$6150	\\   
 1215$-$033 & WHT & 3500$-$7500 & 6000$-$6100	\\  
 1330+022 & M4M & 3890$-$6950 &		\\  
 2141+279 & M4M & 3890$-$6950 &		\\

\hline

\end{tabular}

\end{table*}

$R-K$ colour was fitted in both cases with a typical error of a few
percent.
The observed $R-K$ colours for the host galaxies are obtained from 
{\sc ukirt\,} and \h images (McLure \etal, 1999, Dunlop \etal, 2000), 
and define the basic shape of the host galaxy SED out to $\lambda \simeq 
2 \mu m$. The composite model spectra were appropriately red-shifted before 
calculating the colour, so that they could be compared to the observed 
colours without introducing uncertainties in k-correction. The $R$ band was 
simulated using the filter function, including system response and CCD 
quantum efficiency, for the \h \wf F675W filter, and the K band was 
reproduced using the filter data for the \irc K Ocli filter at 77K combined 
with Mauna Kea atmosphere. 

\section{Results}

The plots showing fits to individual spectra and \xs 
as a function of fitted age are given in Appendix A. The plots for the two-component fit are presented in Figure A1, and those for the two stellar-component plus nuclear contribution are in Figure A2.  
The results for each object are summarized below, under their IAU names, 
with alternative names given in parentheses.
Objects are listed in order of increasing right ascension, within each AGN 
class (radio loud quasars, radio quiet quasars and radio galaxies). 
The telescopes with which the spectra were obtained are also noted; 
M4M denotes the Mayall 4m Telescope at Kitt Peak, and WHT denotes the 4.2m 
William Herschel Telescope on La Palma. Where there are two spectra, the 
first plot is for the spectrum observed with the Mayall 4m Telescope, and 
the second is for the spectrum taken with the William Herschel Telescope.

\subsection{Notes on individual objects}

\subsubsection{Radio loud quasars}

{\bf 0137+012} (L1093) M4M\\
The models give a good fit at 13 Gyr, which is clearly improved by the 
inclusion of a small percentage (0.25\%) of young stars. There is no significant nuclear contribution to the spectrum.
HST imaging has shown that this host galaxy is a large elliptical, with a
half-light radius $r_e = 13$ kpc (McLure et al. 1999).\\
{\bf 0736+017} (S0736+01, OI061) M4M,WHT\\
0736+017 has been observed with both telescopes, and fits to the 
two observed spectra are in good agreement. Both indicate an age of 12 Gyr. 
The M4M spectrum requires a somewhat larger young blue population
(0.75\%) than the WHT spectrum (0.125\%). This may be due to poorer seeing 
at Kitt Peak leading to slightly more nuclear contamination of the slit,
or to the use of slightly different slit positions at the two telescopes.
However, this difference between the observed spectra shortward of 
4000\AA\ leaves the basic form of $\chi^2$ versus age, and the best-fitting 
age of 12 Gyr unaffected. Inclusion of a nuclear component gives a much better fit to the blue end of the M4M spectrum, without changing the age estimation. The size of the fitted young populations are in much better agreement in this case. 
HST imaging has shown that morphologically this host galaxy is a large elliptical, with 
a half-light radius $r_e = 13$ kpc (McLure et al. 1999).\\
{\bf 1004+130} (S1004+13, OL107.7, 4C13.41) WHT\\
The spectrum of this luminous quasar certainly appears to display significant nuclear contamination below the 4000\ang\ break. As a result
a relatively large young population is required to attempt (not
completely successfully) to reproduce the blue end of the spectrum . 
However, the models predict that the underlying stellar population is old 
(12 Gyr). Allowing a nuclear component  to be fit reproduces the blue end of the spectrum much more successfully, without changing the best-fit age estimation.
HST $R$-band imaging indicates the morphology of the host galaxy is dominated by a large
($r_e = 8$ kpc) spheroidal component, but subtraction of this best-fit
model reveals two spiral-arm-type features on either side of the nucleus
(McLure et al. 1999), which may be associated with the young stellar
component required to explain the spectrum.\\
{\bf 1020$-$103} (S1020$-$103, OL133) M4M\\
This object has the second bluest $R-K$ of this sample, which leads to a 
much younger inferred age than the majority of the rest of the sample
(5 Gyr), despite the presence of a rather clear 4000\AA\ break in the
optical spectrum. Ages greater than $\simeq$10 Gyr are rejected by Jimenez' 
models, primarily on the basis of $R-K$ colour.
HST imaging has shown that this host has an elliptical morphology, and a
half-light radius of $r_e = 7$ kpc. (Dunlop et al. 2000).\\
{\bf 1217+023} (S1217+02, UM492) WHT\\
Nuclear contamination can again be seen bluewards of 4000\ang, 
with a correspondingly large young population prediction for the purely stellar population model, which still
fails to account for the very steep rise towards 3000\AA. Hence, a large nuclear contribution is required to reproduce the blue end of the spectrum. The fit achieved 
by the models suggests that the dominant population is old, with a best-fit 
age of 12 Gyr.
HST imaging has shown that this host has an elliptical morphology, and a
half-light radius of $r_e = 11$ kpc. (Dunlop et al. 2000).\\
{\bf 2135+147} (S2135$-$14, PHL1657) WHT\\
2135+147 has a very noisy spectrum, but a constrained fit has still been 
achieved, and an old population is preferred. 2135+147 requires a large $\alpha$, even when a nuclear contribution is fitted.
HST imaging has shown that this host has an elliptical morphology, and a
half-light radius of $r_e = 12$ kpc. (Dunlop et al. 2000).\\
{\bf 2141+175} (OX169) WHT\\
This is another noisy spectrum, which has a relatively large quasar light contribution. An old population is again indicated by the model
fits.
>From optical and infrared imaging this object is known to be complex, but
HST images indicate that it is dominated by a moderate sized ($r_e = 4$
kpc) elliptical component
(see McLure et al. (1999) for further details).\\
{\bf 2247+140} (PKS2247+14, 4C14.82) M4M,WHT\\
2247+140 has been observed with both telescopes. The model fitting indicates 
an old population is required by both spectra - although the two 
observations do not agree precisely, the general level of agreement is
very good, the two $\chi^2$ plots have a very similar form,
and the difference in \xs\ between the alternative 
best-fitting ages of 8 Gyr and 12 Gyr is very small. No significant nuclear contribution to the flux is present.
HST imaging has shown that this host has an elliptical morphology, and a
half-light radius of $r_e = 14$ kpc (Dunlop et al. 2000).\\
{\bf 2349$-$014} (PKS2349$-$01, PB5564) WHT\\
This is a very good fit to a good-quality spectrum, 
showing an obvious improvement when the low-level young population is added. 
Jimenez' models clearly predict that the dominant population is old, 
with  a well-constrained age of 12 Gyr. A very small nuclear contribution 
($\eta$ = 0.050) does not significantly change the results.
HST imaging of this object strongly suggests that it is involved in a
major interaction, with a massive tidal tail extending to the north of
the galaxy. However, the dominant morphological component is a spheroid
with a half-light radius of $r_e = 18$ kpc (McLure et al. 1999).\\

\subsubsection{Radio quiet quasars}

{\bf 0054+144} (PHL909) M4M,WHT\\
There is evidence of relatively large contamination from nuclear emission in 
the spectrum of this luminous quasar taken on both telescopes, and 
the age is not well-constrained, although the fit to the WHT spectrum derived 
from the models again suggests an old age. The $\chi^2$ plots serve to
emphasize how similar the two spectra of this object actually are (as
also discussed by Hughes \etal 2000). Inclusion of a nuclear component in the model better constrains the age and improves the goodness of the fit.
HST imaging of this object has shown that, morphologically, it is
undoubtedly an elliptical galaxy, with a half-light radius $r_e = 8$ kpc
(McLure et al. 1999).\\
{\bf 0157+001} (Mrk 1014) M4M,WHT\\
The age inferred from both the M4M and WHT spectrum of 0157+001 is again 
12 Gyr. The apparently more nuclear-contaminated WHT spectrum does not give 
such a good fit, but 0157+001 is a complex object known to have extended
regions of nebular emission, and the slit positions used for the two
observations were not identical (Hughes \etal 2000). The age is much better
constrained from the more passive M4M spectrum, to which Jimenez' models 
provide a very good fit. Again, it seems that the nuclear contamination does 
not have a great influence on the predicted age of the old population, 
although the fit to the WHT spectrum is greatly improved by including a 
nuclear component in the model. Despite its apparent complexity in both
ground-based and HST images, this host galaxy does again seem to be
dominated by a large spheroidal component, of half-light radius $r_e =
8$ kpc (McLure et al. 1999).\\
{\bf 0204+292} (3C59) WHT\\
Jimenez' models fit the spectrum of this object well, indicating an old underlying stellar 
population ($> 6$Gyr), with a best-fit age of 13 Gyr for the stellar population plus nuclear component model, and a very small young population.
HST imaging has shown this galaxy to be an elliptical, with half-light
radius $r_e = 9$ kpc (Dunlop et al. 2000).\\
{\bf 0244+194} (MS 02448+19) WHT\\
The colour derived from the optical and infrared imaging of this host
galaxy is rather blue ($(R-K)_{obs}$ = 2.34) and this in part leads to a 
fairly young (5 Gyr) age prediction. However, the spectrum is very noisy, 
and, as indicated by the very flat $\chi^2$ plot, the age is not 
strongly constrained. No nuclear flux contamination is fitted.
HST imaging has shown this galaxy to have an elliptical morphology, with half-light
radius $r_e = 9$ kpc (McLure et al. 1999).\\
{\bf 0923+201} WHT\\
The spectrum of 0923+201 is noisy, and it also appears to have some nuclear 
contamination. The fit is therefore improved by inclusion of a nuclear 
component. An old age is strongly preferred
by the form of the \xs\ / age plot, with a best-fit value of 12 Gyr.
HST imaging has shown this galaxy to have an elliptical morphology, with half-light
radius $r_e = 8$ kpc (McLure et al. 1999).\\
{\bf 1549+203} (1E15498+203, LB906, MS 15498+20) WHT\\
This is a good fit, which is clearly improved by the addition of the younger 
population. The slope of the \xs\ / age plot strongly indicates 
an old dominant population, with a best-fit age of 12 Gyr. There is very little evidence of nuclear contamination.
HST imaging has shown this galaxy to be a moderate-sized elliptical
$r_e = 5$ kpc (Dunlop et al. 2000).\\
{\bf 1635+119} (MC1635+119, MC2) WHT\\
This is another very successful fit. An old age (12 Gyr) is inferred.
Again, HST imaging has shown this galaxy to be a moderate-sized elliptical
$r_e = 6$ kpc (McLure et al. 1999).\\
{\bf 2215$-$037} (MS 22152$-$03, EX2215$-$037) WHT\\
2215$-$037 has a noisy spectrum, to which an acceptable fit has nevertheless been 
possible. Jimenez' models suggest an old population, with a best-fitting age 
of 14 Gyr, but this age is not well constrained. This is the only object 
where the inclusion of a nuclear contribution to the fitted spectrum substantially 
changes the age predicted of the dominant stellar population. However, the 
$\chi^2$ plots are very flat after an age of 5 Gyr, and the age is not 
strongly constrained in either case.
HST imaging has shown this galaxy to have an elliptical morphology, with
$r_e = 7$ kpc (Dunlop et al. 2000).\\
{\bf 2344+184} (E2344+184) M4M,WHT\\
2344+184 has been observed with both telescopes, and the fits to both 
spectra are in good agreement $-$ although, formally, 
two different ages are predicted. This is because 
the \xs\ / age plots are fairly flat after about 8 Gyr.
The fits to both observations suggest an old dominant population, with a 
small young blue population improving the fit. There is no 
significant change in the predictions when a nuclear flux component is 
included in the model.
HST imaging has shown this to be one of the few host galaxies in the
current sample to be disc-dominated. However, the nuclear component is in
fact sufficiently weak that this object should really be classified as a
Seyfert galaxy rather than an RQQ (McLure et al. 1999).\\

\subsubsection{Radio galaxies}
\begin{table}

\caption{Rest frame emission lines masked out in the \xs\ fit.}

\begin{tabular}{cl}

	\hline
	Masked region / \ang & Emission line			\\
	\hline
	3720 $-$ 3735 & OII 3727 					\\

	3860 $-$ 3880 & NeIII 3869  				\\

	4840 $-$ 5020 & OIII 4959, OIII 5007, H$_{\beta}$ 4861 	\\

	\hline

\end{tabular}

\end{table}

{\bf 0230$-$027} (PKS0230$-$027) WHT\\
0230$-$027 has a very noisy spectrum, and the colour of the host as
derived from optical and infrared imaging is very blue 
($(R-K)_{obs}$ = 2.09). Consequently the best-fitting age derived using 
the models is 1 Gyr, with no younger component, but it is clear that
little reliance can be placed on the accuracy of this result. Allowing for a contribution from quasar light does not improve the fit.
HST imaging has shown this galaxy to have an elliptical morphology, with
$r_e = 8$ kpc (Dunlop et al. 2000).\\
{\bf 0345+337} (3C93.1) WHT\\
0345+337 requires no young component or nuclear flux contribution at all, and an old age, of 12 Gyr, 
is clearly indicated by the models.
HST imaging has shown this galaxy to be a large elliptical, with
$r_e = 13$ kpc (McLure et al. 1999).\\
{\bf 0917+459} (3C219, 3C219.0) WHT\\
This is an excellent fit, with an old age produced by the models (12 Gyr), 
together with a very small young population.
HST imaging has shown this galaxy to be a large elliptical, with
$r_e = 11$ kpc (McLure et al. 1999).\\
{\bf 1215$-$033} WHT\\
Jimenez' models suggest that the population of 1215$-$033 is universally 
old (best-fit age, 13 Gyr), with no young component or nuclear contribution required.
HST imaging has shown this galaxy to be a large elliptical, with
$r_e = 9$ kpc (Dunlop et al. 2000).\\
{\bf 1330+022} (3C287.1) M4M\\
An excellent fit to the data is produced by Jimenez' models, 
indicating that the dominant population is old, with a best-fit age of 8 Gyr.
HST imaging has shown this galaxy to be a large elliptical, with
$r_e = 16$ kpc (Dunlop et al. 2000).\\
{\bf 2141+279} (3C436) M4M,WHT\\
This is another very successful, and well-constrained fit,
with an inferred age of 12 Gyr.
HST imaging has shown this galaxy to be a large elliptical, with
$r_e = 21$ kpc (McLure et al. 1999).\\

\subsection{Sample overview}

The results illustrated in Appendix A are summarised in Table 3. 
It should be noted that the 4000\ang\ break typical of evolved stellar 
populations is present in the majority of the observed spectra
(see Appendix A and Hughes \etal 2000), so we can be 
confident in fitting stellar population models to the data. 
The plots clearly show that the addition of even a very small amount of 
secondary star formation to the simple, near-instantaneous star-burst models 
reproduces the blue end of the observed host galaxy spectra much more 
successfully (and in most cases very well) 
than does a single stellar population. Including a nuclear component further improves the fit to the blue end, especially for those spectra not well fit by purely stellar light. At the same time, the red end of the 
spectra, plus the observed $R-K$ colours generally require that the underlying 
stellar populations are old.

\begin{table*}

\centerline{\epsfig{file=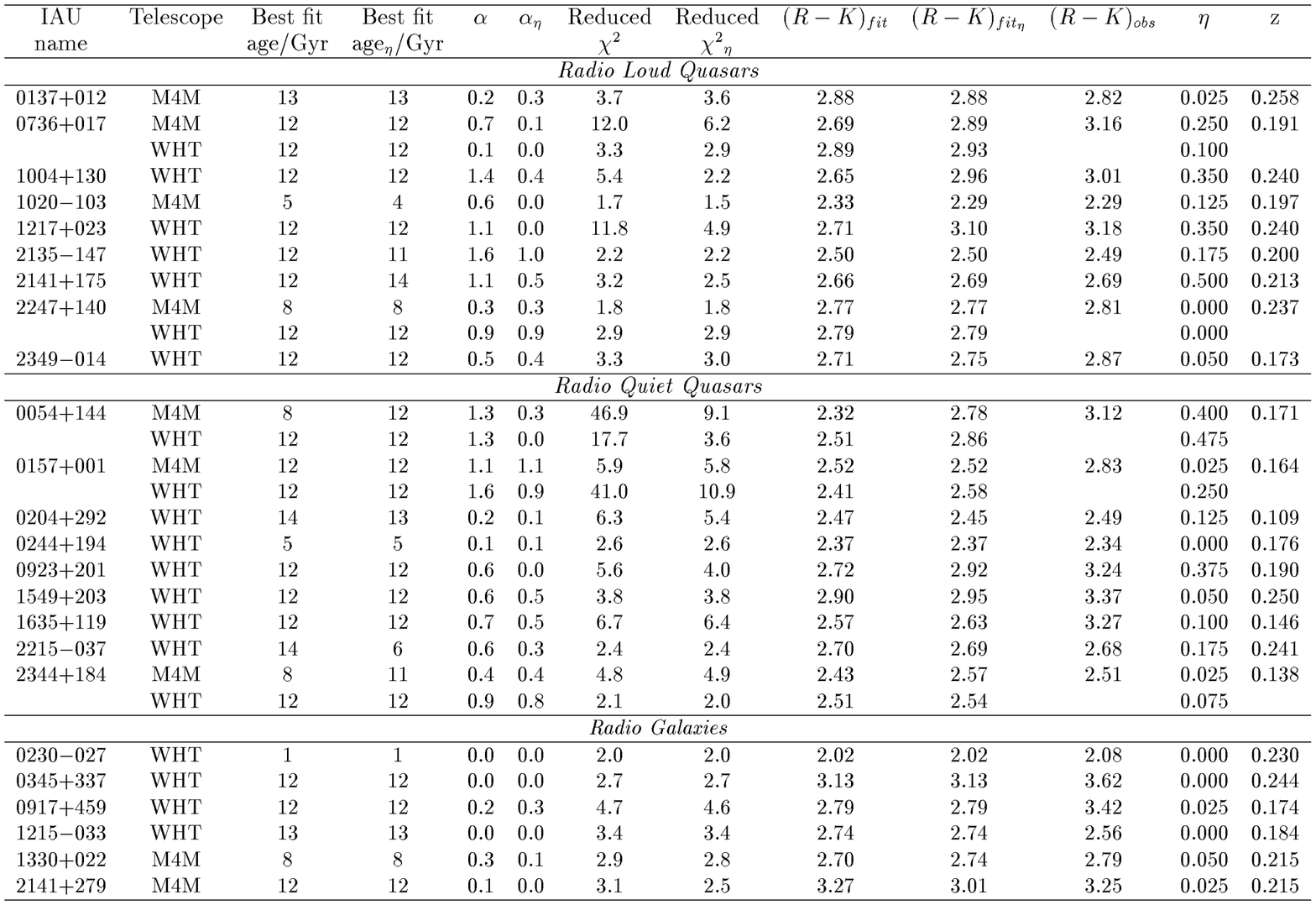,width=18cm,angle=0,clip=}}

\setcounter{table}{2}

\caption{Results from the simultaneous fitting to the AGN host sample of the two-component model spectra (using the solar metallicity models of Jimenez et al. 2000) and $R-K$ colour. $\alpha$ is the percentage young population, by mass. Results are also p
resented for the fits including the subtraction of a nuclear component from the observed spectrum. $\eta$ is the fraction of nuclear flux subtracted, and the corresponding results are denoted by the subscript $\eta$.} \label{tbl-3}

\end{table*}

The most meaningful output from the model fitting is a constraint
on the minimum age of the host galaxies; 
the \xs\ plots show that often the best-fit age is not strongly constrained 
but the trend is clearly towards old ($\geq$ 8 Gyr) stellar populations. In general, young populations are strongly excluded.

The peaks and troughs in \xs\ as a function of model age 
appear to be the result of real features of the population evolution
synthesis, rather than being due to, for example, poor sampling
(Jimenez, private communication).

\begin{figure*}
\centerline{\epsfig{file=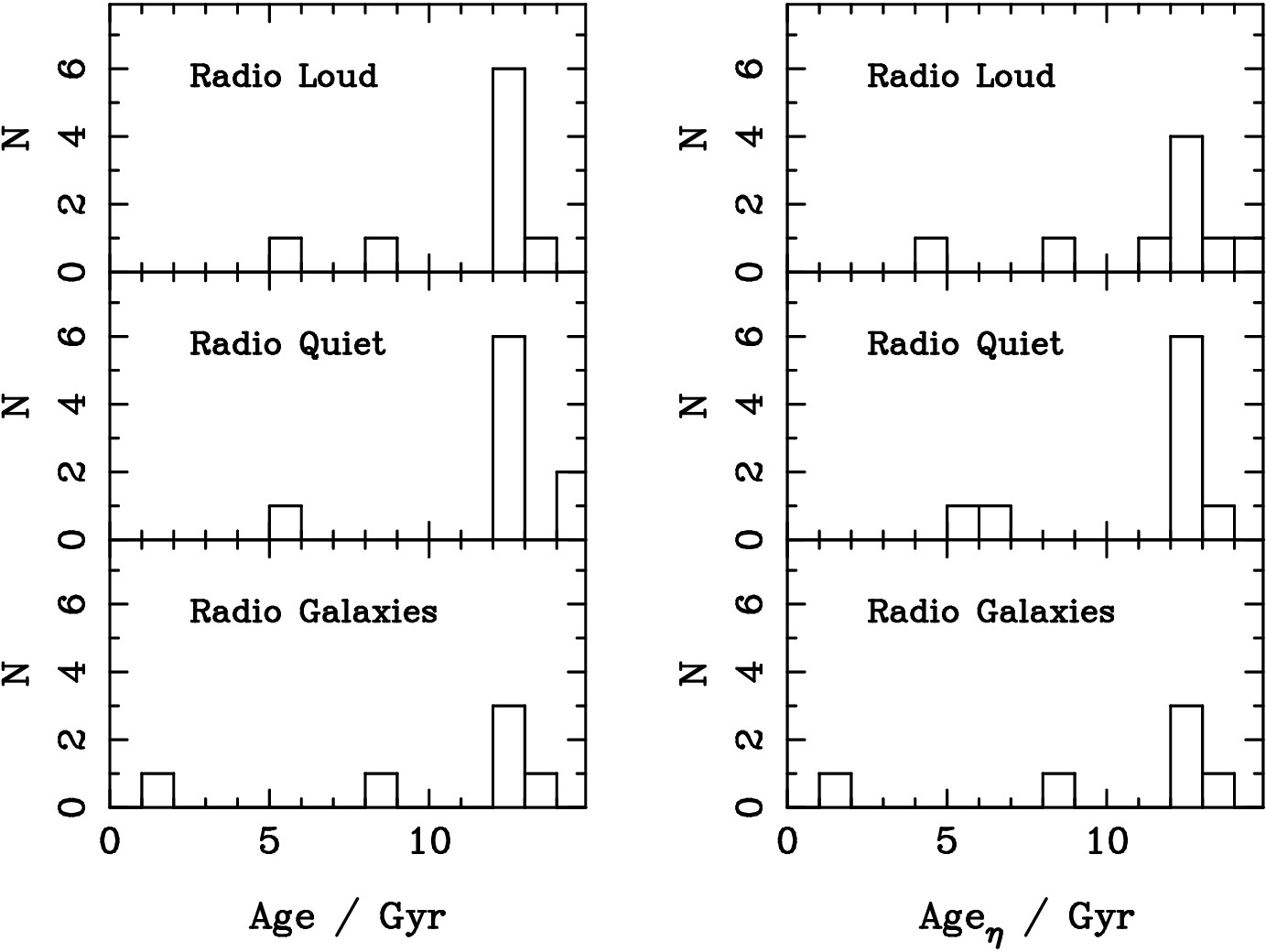,width=9cm,angle=-90,clip=}}
        \caption{The age distribution of the dominant stellar populations of the sample host galaxies. These are the best-fitting results from Jimenez' solar metallicity models. On the left, results are shown for the fits to the data using a two-component
 model. The results on the right are those for the two-component model plus a nuclear contribution. The populations are predominantly old (12-14 Gyr) in both cases.}
	\vspace{0.3in}
\centerline{\epsfig{file=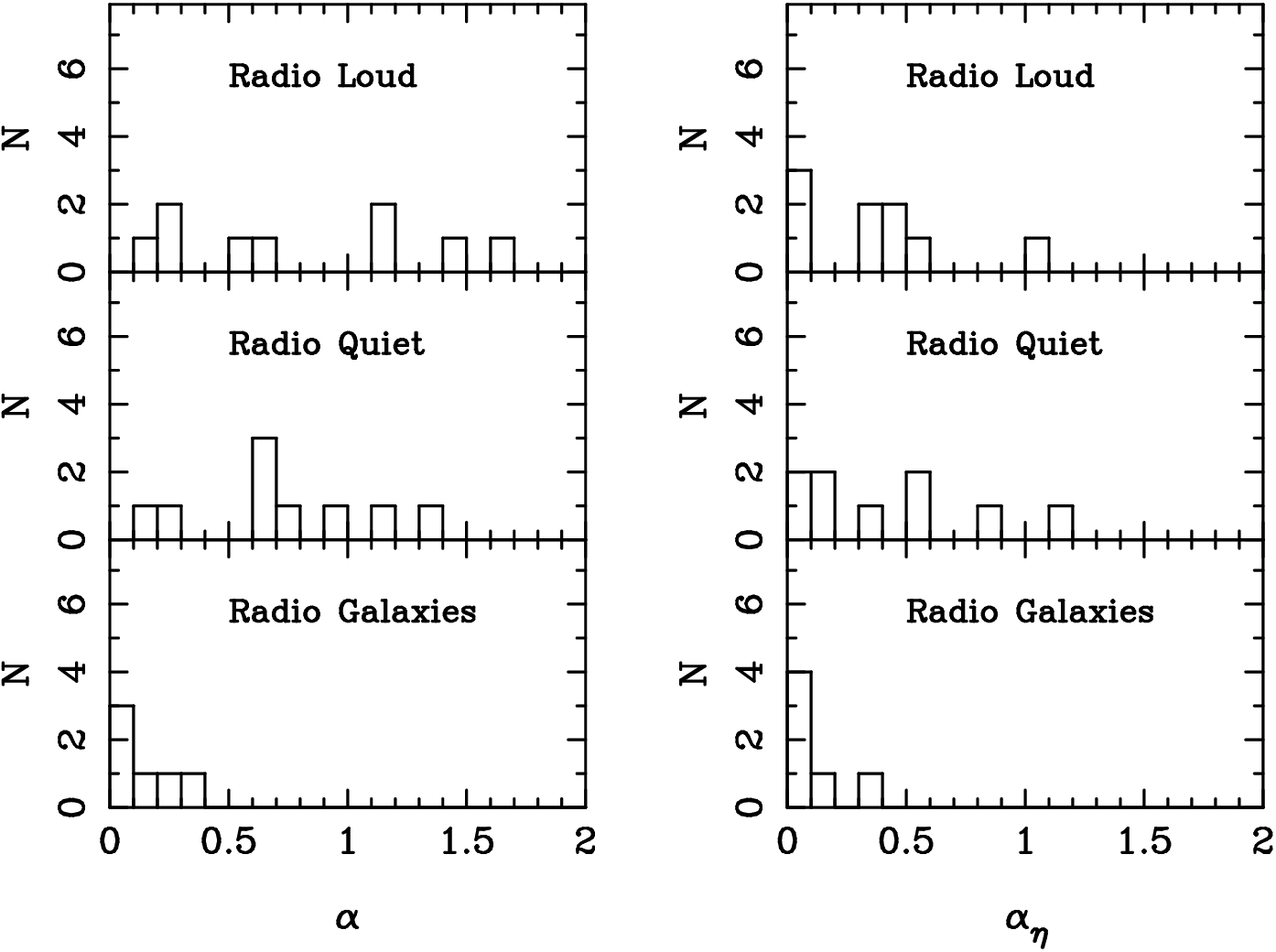,width=9cm,angle=-90,clip=}}
        \caption{The distribution of $\alpha$, the percentage 
contribution (by mass) of the 0.1 Gyr component. 
Where results have been obtained from two spectra for one object, the 
best-fitting result has been adopted. Again, the results on the left are for 
fits using the two-component model, and those on the right are for fits with the 
two-component model plus a nuclear contribution. Allowing for the
possibility of a nuclear contribution means that a 
smaller $\alpha$ is required to fit the blue end of the spectrum, and the 
apparent difference between the radio galaxies and radio-loud and quiet 
quasars is reduced to a statistically insignificant level.}

\end{figure*}

\section{Discussion}

Figure 1 shows the distribution of best-fit ages estimated using Jimenez' 
models. The left-hand panel shows the ages of the dominant stellar
component which result from fitting the two-component (stars only) model, while the right-hand
panel shows the ages of the dominant stellar component in the case of the 
3 component model (i.e. two stellar components + a contribution from 
scattered nuclear light). The host galaxies of the AGN in each sample 
are predominantly old, and this result is unaffected by whether or not
one chooses to include some nuclear contribution to assist in fitting the
blue end of the spectrum. 
This general result also appears to be relatively unaffected by the 
precise choice 
of stellar population model;
the models of Bruzual and Charlot (2000), fitted to the host galaxies 
with the same process, reproduce the results for Jimenez' models to 
within a typical accuracy of 1-3 Gyr. We have not attempted to fit the stellar population models of Yi et al. (2000), because of problems we have found in their MS rate of evolution (Nolan et al., 2000).

Inclusion of a nuclear contribution in the 3-component models obviously
raises the question of whether a young stellar population component
is really necessary at all. We thus also explored the results of 
fitting a two-component model consisting of a single stellar population 
plus nuclear contribution. Such a model  adequately reproduces the 
two-population-plus-nuclear-component results 
for the {\it redder} galaxies, but in fact the 
spectra of the bluest galaxies cannot be adequately reproduced  by these 
models; the resulting serious increase in minimum \xs\ demonstrates 
that inclusion of a young population component is necessary 
to achieve a statistically acceptable fit to these data.

Figure 2 shows the distribution of percentage contribution (by mass),
$\alpha$, of
the young (0.1 Gyr) component to the spectra of the hosts.
The left-hand panel shows the values of $\alpha$ as derived from 
fitting the two-component (stars only) model, while the right-hand
panel shows the values of $\alpha$ produced by the 
3 component model (i.e. two stellar components + a contribution from 
scattered nuclear light).

Where two spectra of the same object have been obtained with alternative
telescopes/instruments, the derived ages
of the dominant stellar componenets are reproduced 
reassuringly well. There are, however, small discrepancies in the 
estimated percentage of young population present (see Table 3). 
This effect is suggestive that at least some of the blue light might be
due to a scattered nuclear contribution, the strength of which would be
highly dependent on the seeing at the time of observation, and on the
precise repeatability of slit placement relative to the galaxy core.
Interestingly, when a nuclear component is included 
with the stellar flux model, an even smaller percentage of 0.1 Gyr 
stellar population is required to fit the blue end of the spectra, and 
(more importantly) the difference in $\alpha$ between two spectra of the same object is
generally reduced. This provides further support for the suggestion that
some of the bluest quasar host spectra remain contaminated by quasar
light at the shortest wavelengths, and indicates that the right-hand
panel of Figure 2 provides a more realistic estimate of the level of
on-going star-formation in the host galaxies. While this figure still
appears to suggest that at least some quasar hosts display higher levels
of ongoing star-fomration activity than do radio galaxies, statistically
this `result' is not significant.

There are three galaxies which have very low age estimates, namely 0230$-$027, 
0244+1944 and 1020$-$103. These objects have the bluest observed 
$R-K$ colours, so it may be expected that the fitted ages would be 
younger than the rest 
of the sample, and that these populations are genuinely young. As 
discussed above, it may be that these objects are bluer because of scattered 
nuclear light contaminating the host galaxy spectrum. However, it seems 
unlikely that their fitted ages are low simply because of this, 
because elsewhere in our sample, where 
two spectra have been obtained of the same object, the amount of nuclear 
contamination present does not significantly affect the age estimation 
({\em e.g.} 0736+017 and 0157+001). Moreover, the inclusion of a nuclear 
component to the fit does not change the estimated age distribution of the 
host galaxies. If these ages are in error, then a more likely 
explanation, supported by the relative
compactness of these particular host galaxies, is that nuclear and host
contributions have been imperfectly separated in the $K$-band images,
leading to an under-estimate of the near-infrared luminosity of the host.

The result of the three-component fitting process which also allows a contribution from
scattered nuclear light is that there are in fact only 3 host galaxies in 
the sample for which there is evidence that $\alpha > 0.5$. One of these
is the host of a radio-loud quasar (2135$-$147) but, as can be
seen from Fig A2, this spectrum is one of the poorest (along with
0230$-$027, 
the only apparently young radio galaxy) in the entire dataset.
The 2 convincing cases are both the hosts of radio-quiet quasars,
namely 0157$+$001 and 2344$+$184.

Within the somewhat larger sample of 13 RQQs imaged with the HST by McLure
et al. (1999) and Dunlop et al. (2000), 4 objects showed evidence for a
disk component in addition to a bulge, namely 0052+251, 0157+001,
0257+024, and 2344+184. Since we do not possess spectra of 0052+251
and 0257+024 this means that there is a 1:1 correspondence
between the objects which we have identified on the basis of this
spectroscopic study as having recent star-formation activity, and those
which would be highlighted on the basis of HST imaging as possessing a
significant disk component. This straightforward match clearly
provides us with considerable confidence that the spectral decomposition 
attempted here has been effective and robust.
Finally we note that it is almost certainly significant that 0157+001, 
which has the largest starburst component ($\alpha = 1.1$) based on this 
spectroscopic analysis, is also the only IRAS source in the sample.

\section{Conclusions}

We conclude that the hosts of all three major classes of AGN contain 
predominantly old stellar populations ($\simeq 11$ Gyr) by $z \simeq 0.2$.
This agrees well with the results of McLure \etal (1999), and Dunlop
\etal (2000) who compare host 
galaxy morphologies, luminosities, scale lengths and colours in the same 
sample, and conclude that the hosts are, to first order,
indistinguishable from `normal' quiescent giant elliptical galaxies.

The best-fitting age of the dominant stellar population is {\it not} a function
of AGN class. For the purely stellar models, the fitted percentage contribution of the blue component is,
however, greater in the quasar hosts than in the radio galaxies; the median
values are 0.6\% for the 9 radio-loud quasars, 0.6\% for the 9
radio-quiet quasars, and 0.05\% for the 6 radio galaxies. However, when a nuclear component is included, the median
values are 0.3\% for the radio-loud quasars, 0.3\% for the radio-quiet quasars, and 0.00\% for the radio galaxies. Performing a Kolmogorov-Smirnov test on these results 
yields a probability greater than 0.2 that the percentage of young 
stellar population in host galaxies is in fact also not a function of AGN class.

These results strongly support the conclusion that the host galaxies of all 
three major classes of AGN are massive ellipticals, dominated by old stellar 
populations.
    \\
    \\
    \\
{\bf ACKNOWLEDGEMENTS}\\
    \\
Louisa Nolan acknowledges the support provided by the award of a PPARC
Studentship. Marek Kukula and David Hughes acknowledge the support
provided by the award PPARC PDRAs, while Raul Jimenez acknowledges the award 
of a PPARC Advanced Fellowship. We thank an anonymous referee for
perceptive comments which helped to clarify the robustness of our results
and improved the clarity of the paper.
    \\
{\bf REFERENCES}\\
    \\
Bahcall J.N., Kirhakos S., Saxe D.H., Schneider D.P., 1997, ApJ, 479, 642\\
Bruzual A.G., Charlot S., 2000, in preparation\\
Disney M.J. et al., 1995, Nat, 376, 150\\
Dunlop J.S., 2000, In: `The Hy-Redshift Universe: Galaxy Formation and
Evolution at High Redshift', ASP Conf. Ser., Vol 193, eds. A.J. Bunker \&
W.J.M. van Breugel, in press, (astro-ph/9912380)\\
Dunlop J.S., McLure R.J., Kukula M.J., Baum S.A., O'Dea C.P., Hughes D.H., 2000, MNRAS, in press\\ 
Dunlop J.S., Taylor G.L., Hughes D.H., Robson E.I., 1993, MNRAS, 264, 455\\
Hooper E.J., Impey C.D., Foltz C.B., 1997, ApJ, 480, L95\\
Hughes D. H., Kukula M.J., Dunlop J.S., Boroson T., 2000, MNRAS, in press\\
Hutchings J.B., Morris S.C., 1995, AJ, 109, 1541\\
Jimenez R., Dunlop J.S., Peacock J.A., Padoan P., MacDonald J., J$\o$rgensen U.G., 2000, MNRAS, in press\\
Jimenez R., Padoan P., Matteucci F., Heavens A.F., 1998, MNRAS, 299, 123\\
Magorrian J., et al., 1998, AJ, 115, 2285\\
McLure R.J., Kukula M.J., Dunlop J.S., Baum S.A., O'Dea C.P., Hughes D.H., 1999, MNRAS, in press, (astro-ph/9809030)\\
Nolan L.A., Dunlop J.S., Jimenez R., 2000, MNRAS, submitted, (astro-ph/0004325)
Schade, D.J., Boyle, B.J., Letawsky, M., 2000, MNRAS, in press\\
Taylor G.T., Dunlop J.S., Hughes D.H., Robson E.I., 1996, 
MNRAS, 283, 930\\
Yi S., Brown T.M., Heap S., Hubeny I., Landsman W., Lanz T., Sweigart A., 2000, ApJ, submitted\\

\appendix

\section{Spectra and \xs\ plots}
The fits for all the off-nuclear spectra are given. In Fig A1, the rest frame spectra are in the first column (black), with the best-fitting two-component model spectra (Jimenez et al., 2000) superimposed (green). The spectra of the single-aged old popula
tion (red) is given for comparison.  In Fig A2, the fits allowing for an additional contribution to the flux from the nucleus are presented. The key is the same as in A1, with the additional blue line representing the best-fitting two-component model flux
 plus the nuclear flux contribution. 

The second column of plots shows the \xs evolution with age for the dominant older population. The third column shows the best-fit \xs as a function of percentage young population, $\alpha$, for fixed ages of the dominant component. The models have solar 
metallicity. The subscript $\eta$ denotes results obtained by including the nuclear contribution.

Where there are two spectra of the same object, the spectrum given first is the one observed on the Mayall 4m Telescope, and the second is that observed using the William Herschel Telescope.The data for the following objects have been smoothed using a Han
ning function: 2135+147 (RLQ), 2141+175 (RLQ), 0244+194 (RQQ), 0923+201 (RQQ), 1549+203 (RQQ), 2215-037 (RQQ), 0230-027 (RG) and 0345+337 (RG).

\newpage
\begin{figure*}
\centerline{ {\LARGE {\em Radio Loud Quasars}} }
\centerline{\epsfig{file=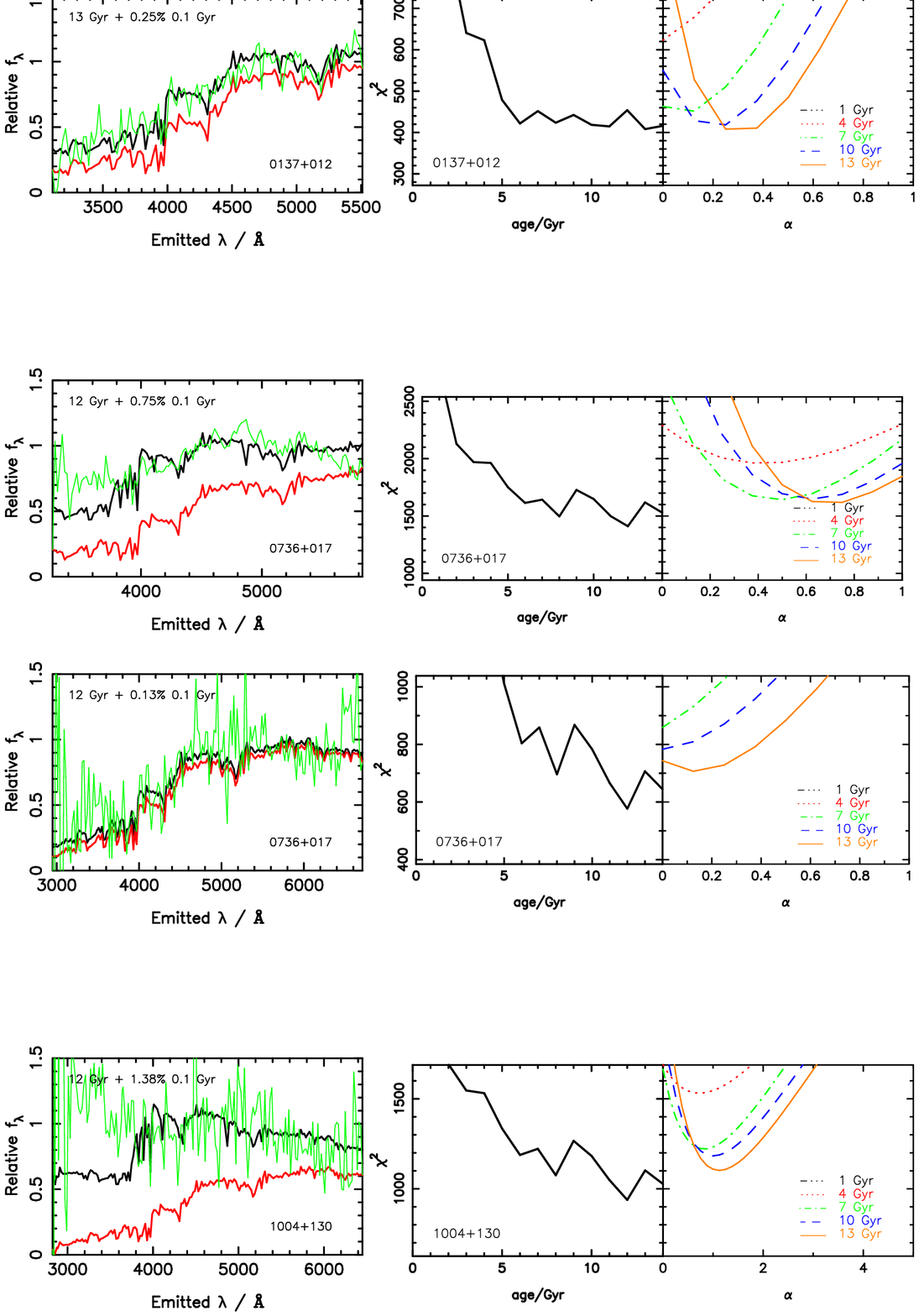,width=15cm,angle=0,clip=}}
	\vspace{-1.5cm}
        \caption{Model fits to the off-nuclear rest frame spectra, for
each object, with the corresponding $\chi^{2}$ plots. The rest-frame 
host-galaxy spectra are in the first column (lightest grey), with the 
best-fitting two-component model spectra (Jimenez et al., 2000) superimposed (black). 
The spectra of the single-aged old population (mid-grey, lowest line) is given 
for comparison. The second column shows the \xs evolution with age for the 
dominant older population and the third column shows the best-fit \xs\ as a 
function of percentage young population, $\alpha$, for fixed ages of the 
dominant component. All models have solar metallicity. Where there are two 
spectra of the same object, the spectrum given first is the one observed on 
the Mayall 4m Telescope, and the second is that observed using the 
William Herschel Telescope.
The data for the following 
objects have been smoothed using a Hanning function: 2135+147 (RLQ), 2141+175 (RLQ), 0244+194 
(RQQ), 0923+201 (RQQ), 1549+203 (RQQ), 2215$-$037 (RQQ), 0230$-$027 (RG) 
and 0345+337 (RG).}
\end{figure*}
 
\begin{figure*}
\setcounter{figure}{0}
\centerline{{\LARGE {\em Radio Loud Quasars}} }
\centerline{\epsfig{file=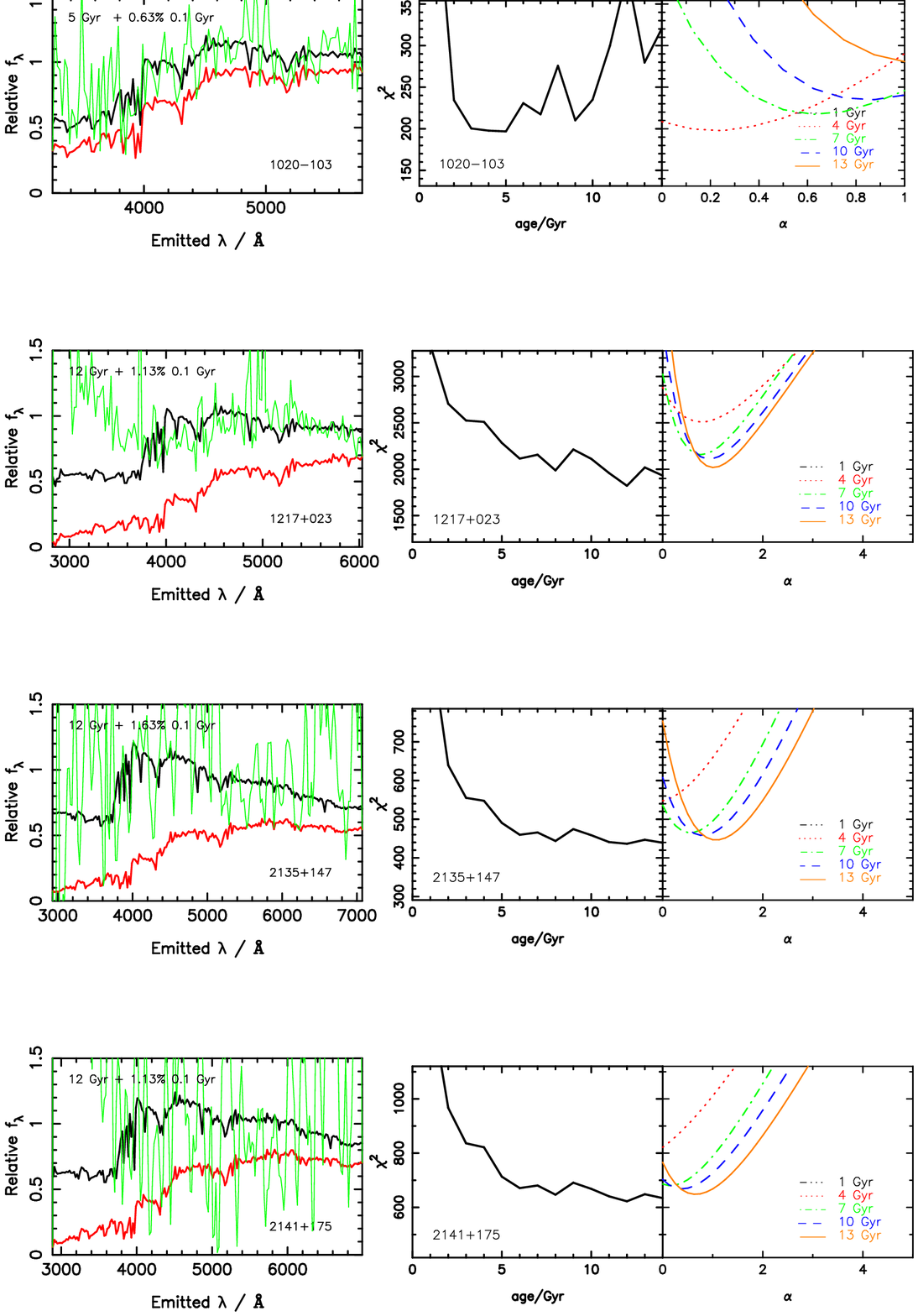,width=16cm,angle=0,clip=}}
	\vspace{-1.5cm}
        \caption{Model fits to the off-nuclear rest frame spectra, continued. }
\end{figure*}

\begin{figure*}
\setcounter{figure}{0}
\centerline{ {\LARGE {\em Radio Loud Quasars}} }
\centerline{\epsfig{file=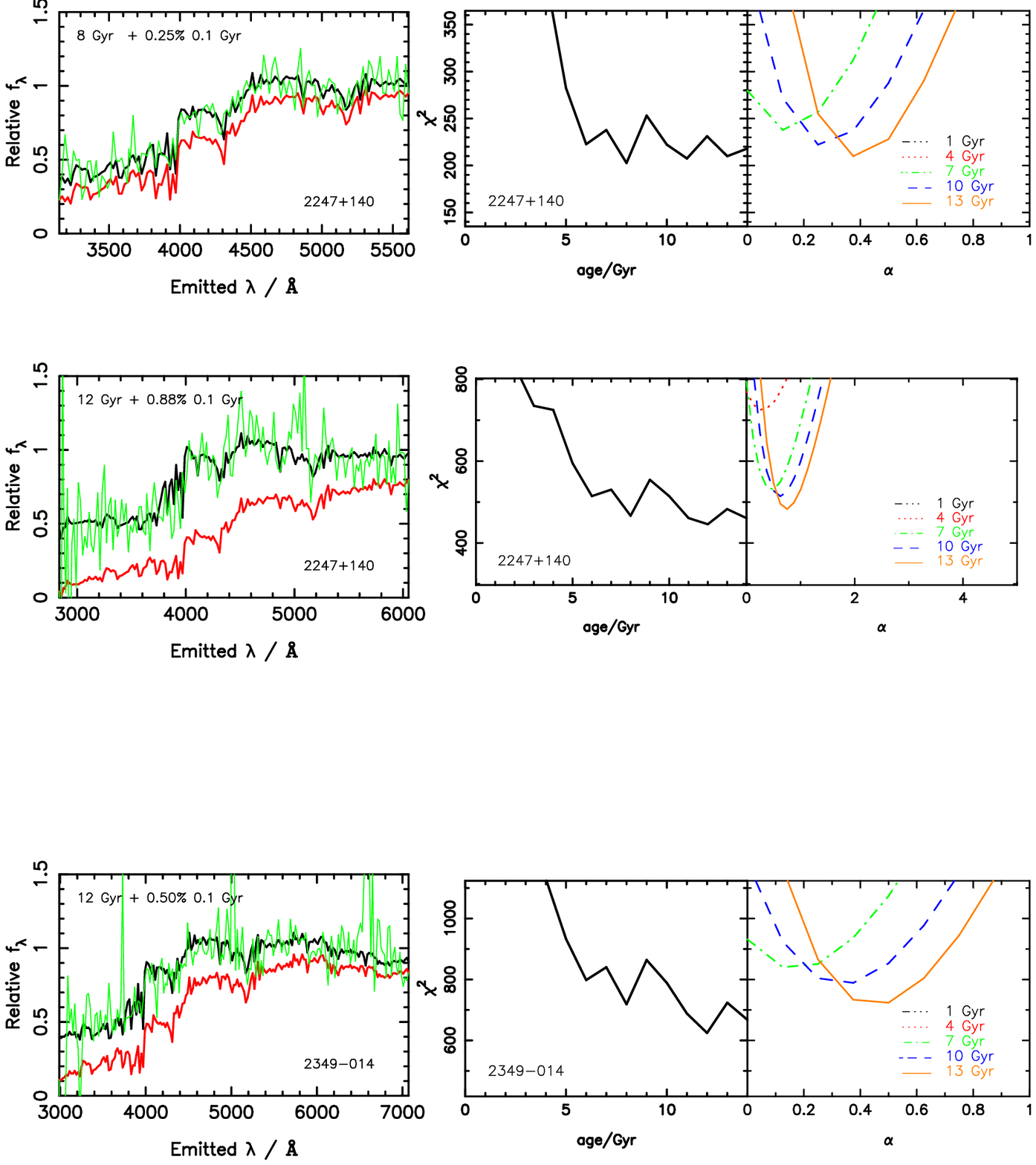,width=16cm,angle=0,clip=}}
	\vspace{-6.0cm}
        \caption{Model fits to the off-nuclear rest frame spectra, continued. }
\end{figure*}

\begin{figure*}
\setcounter{figure}{0}
\centerline{ {\LARGE {\em Radio Quiet Quasars}} }
\centerline{\epsfig{file=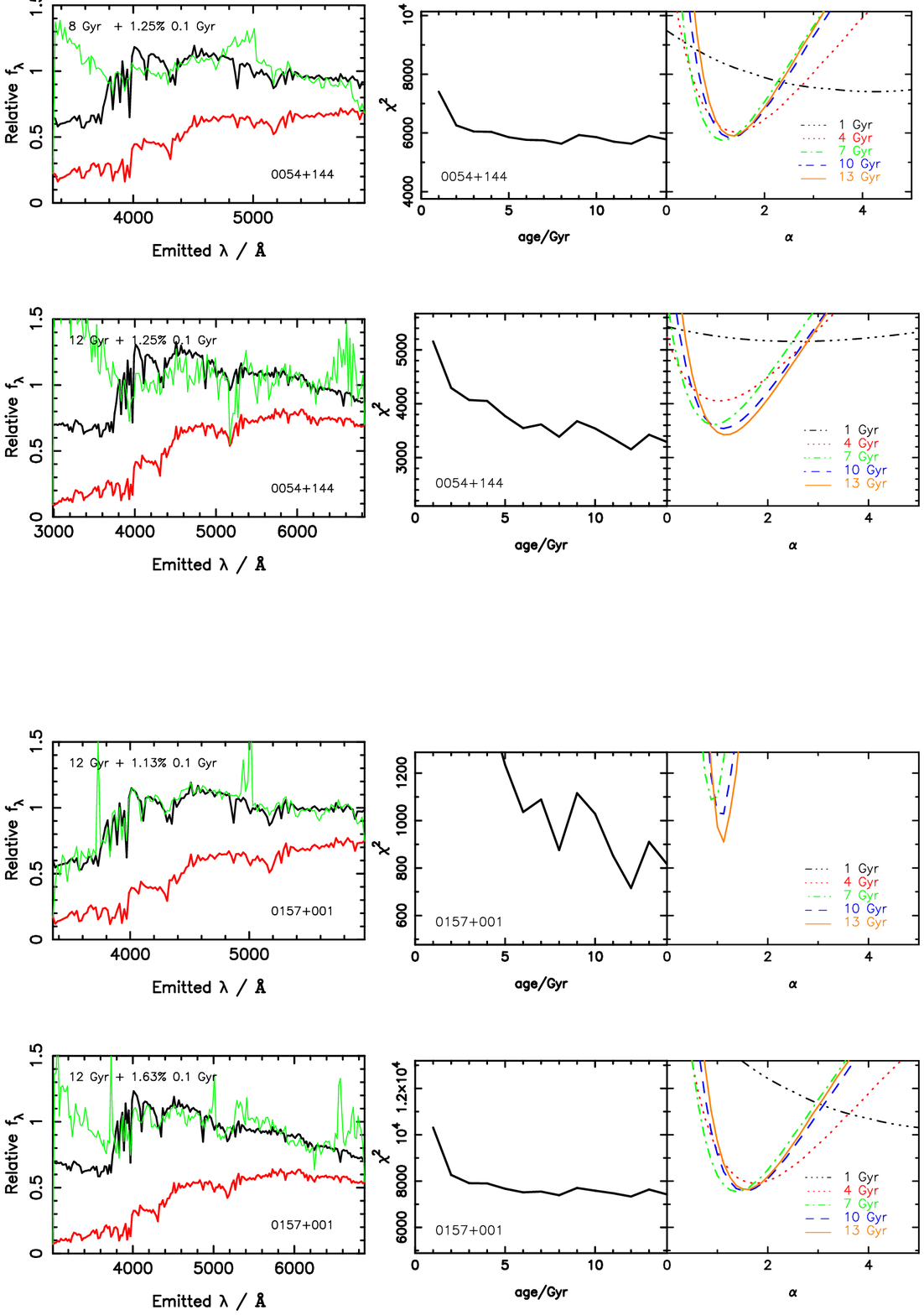,width=16cm,angle=0,clip=}}
	\vspace{-1.5cm}
        \caption{Model fits to the off-nuclear rest frame spectra, continued. }
\end{figure*}

\begin{figure*}
\setcounter{figure}{0}
\centerline{ {\LARGE {\em Radio Quiet Quasars}} }
\centerline{\epsfig{file=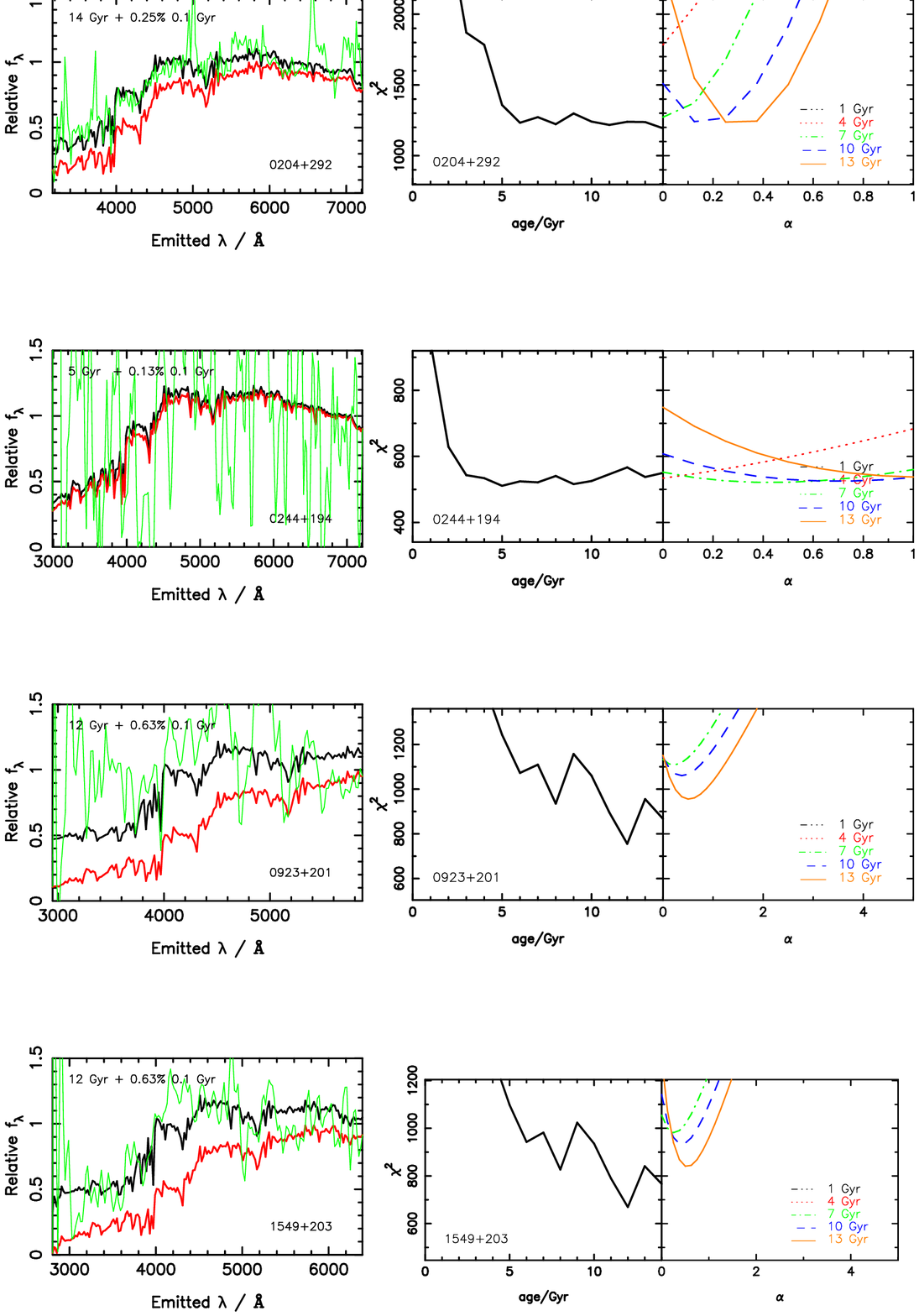,width=16cm,angle=0,clip=}}
	\vspace{-1.5cm}
        \caption{Model fits to the off-nuclear rest frame spectra, continued. }
\end{figure*}

\begin{figure*}
\setcounter{figure}{0}
\centerline{ {\LARGE {\em Radio Quiet Quasars}} }
\centerline{\epsfig{file=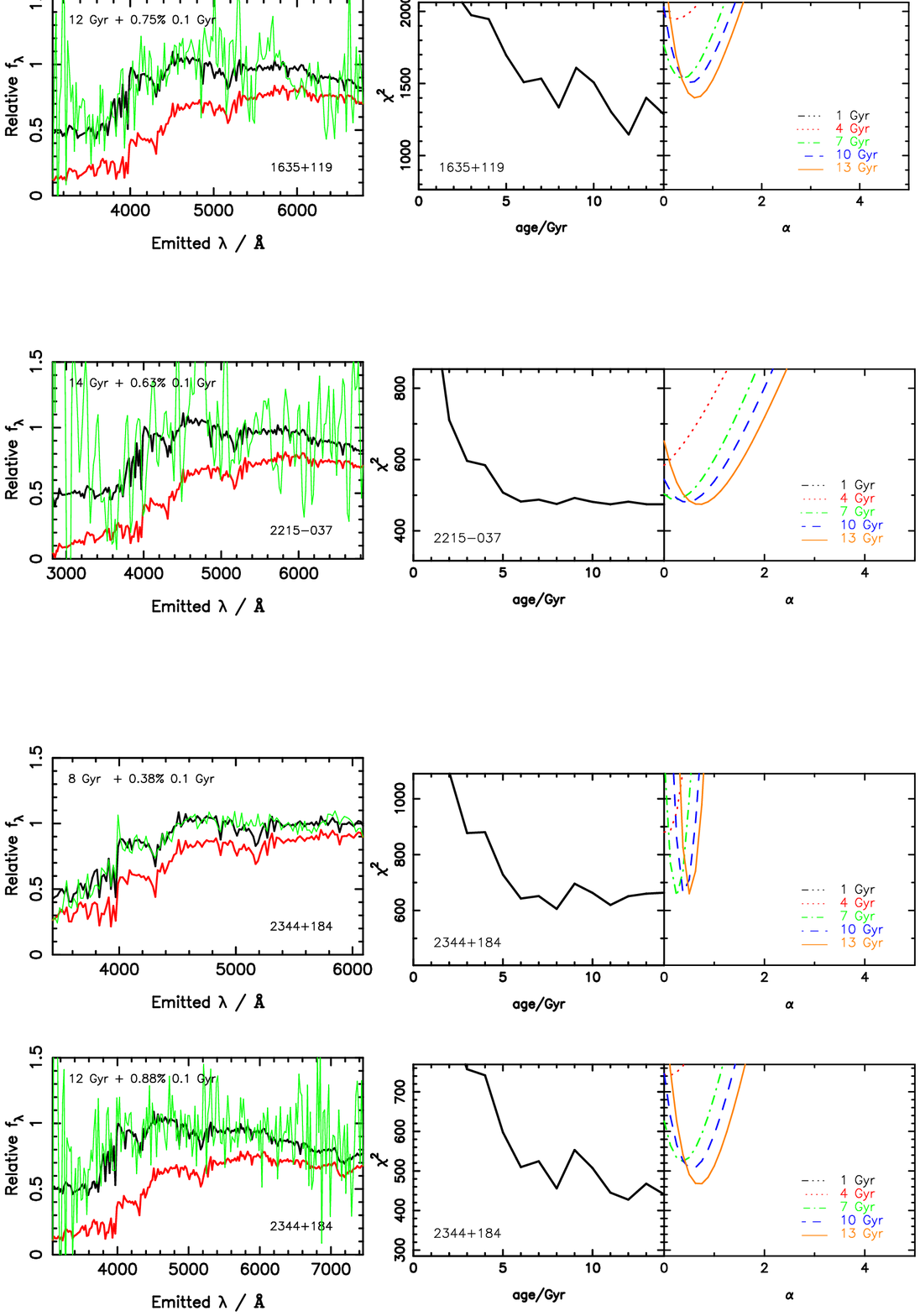,width=16cm,angle=0,clip=}}
	\vspace{-1.5cm}
        \caption{Model fits to the off-nuclear rest frame spectra, continued. }
\end{figure*}

\begin{figure*}
\setcounter{figure}{0}
\centerline{ {\LARGE {\em Radio Galaxies}} }
\centerline{\epsfig{file=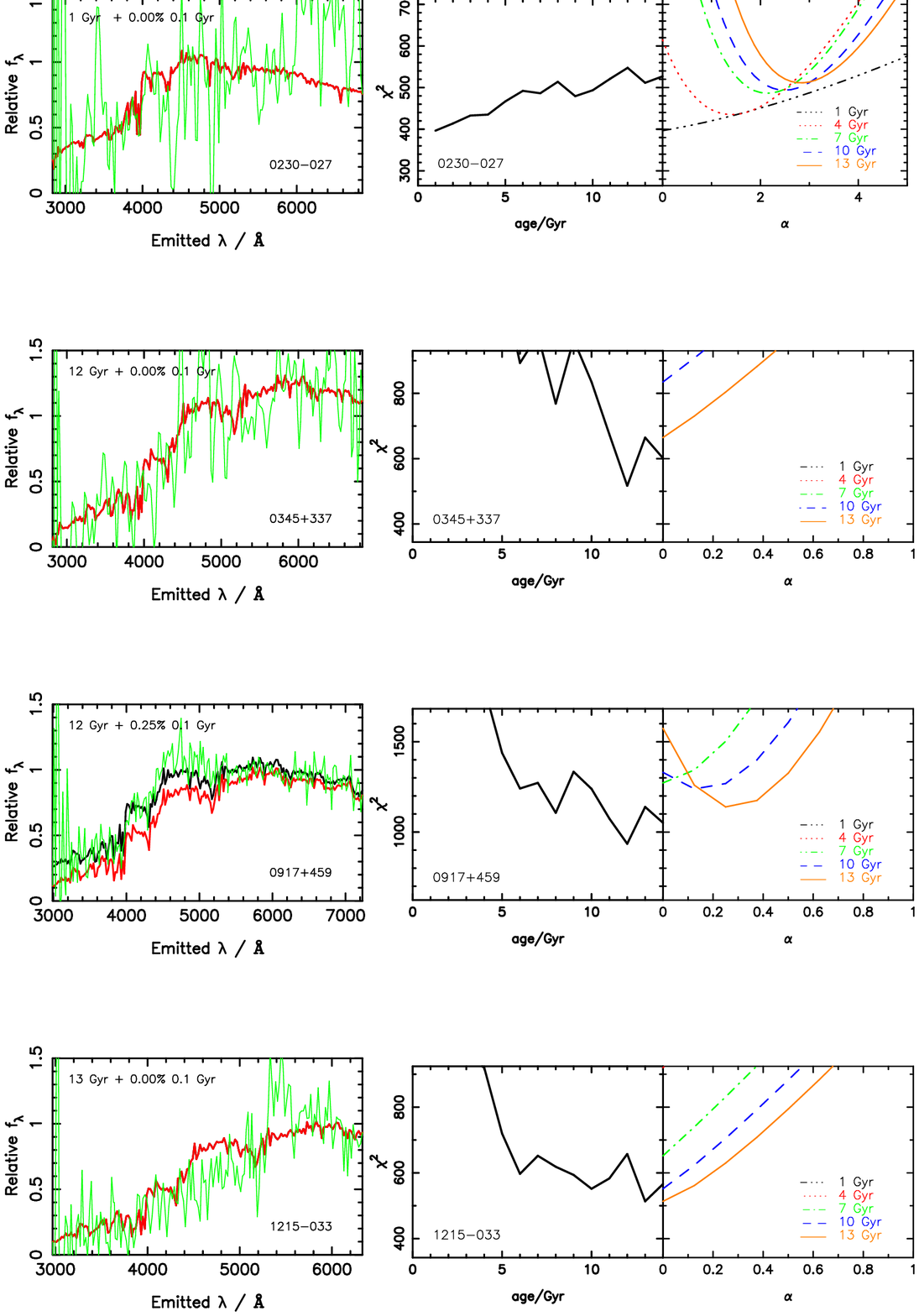,width=16cm,angle=0,clip=}}
	\vspace{-1.5cm}
        \caption{Model fits to the off-nuclear rest frame spectra, continued. }
\end{figure*}

\begin{figure*}
\setcounter{figure}{0}
\centerline{ {\LARGE {\em Radio Galaxies}} }
\centerline{\epsfig{file=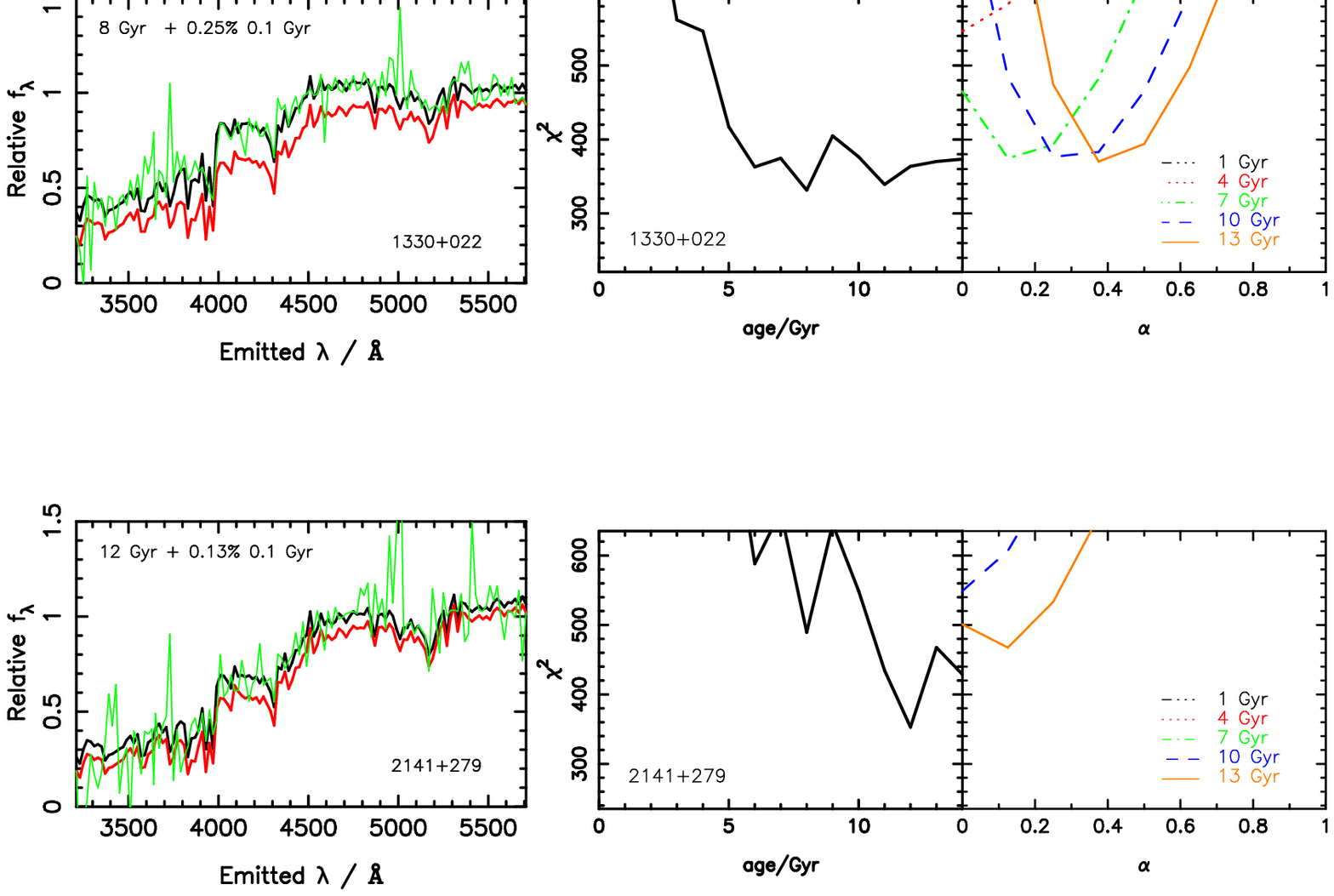,width=16cm,angle=0,clip=}}
	\vspace{-11cm}
        \caption{Model fits to the off-nuclear rest frame spectra, continued. }
\end{figure*}


\begin{figure*}
\centerline{{\LARGE {\em Radio Loud Quasars}} }
\centerline{\epsfig{file=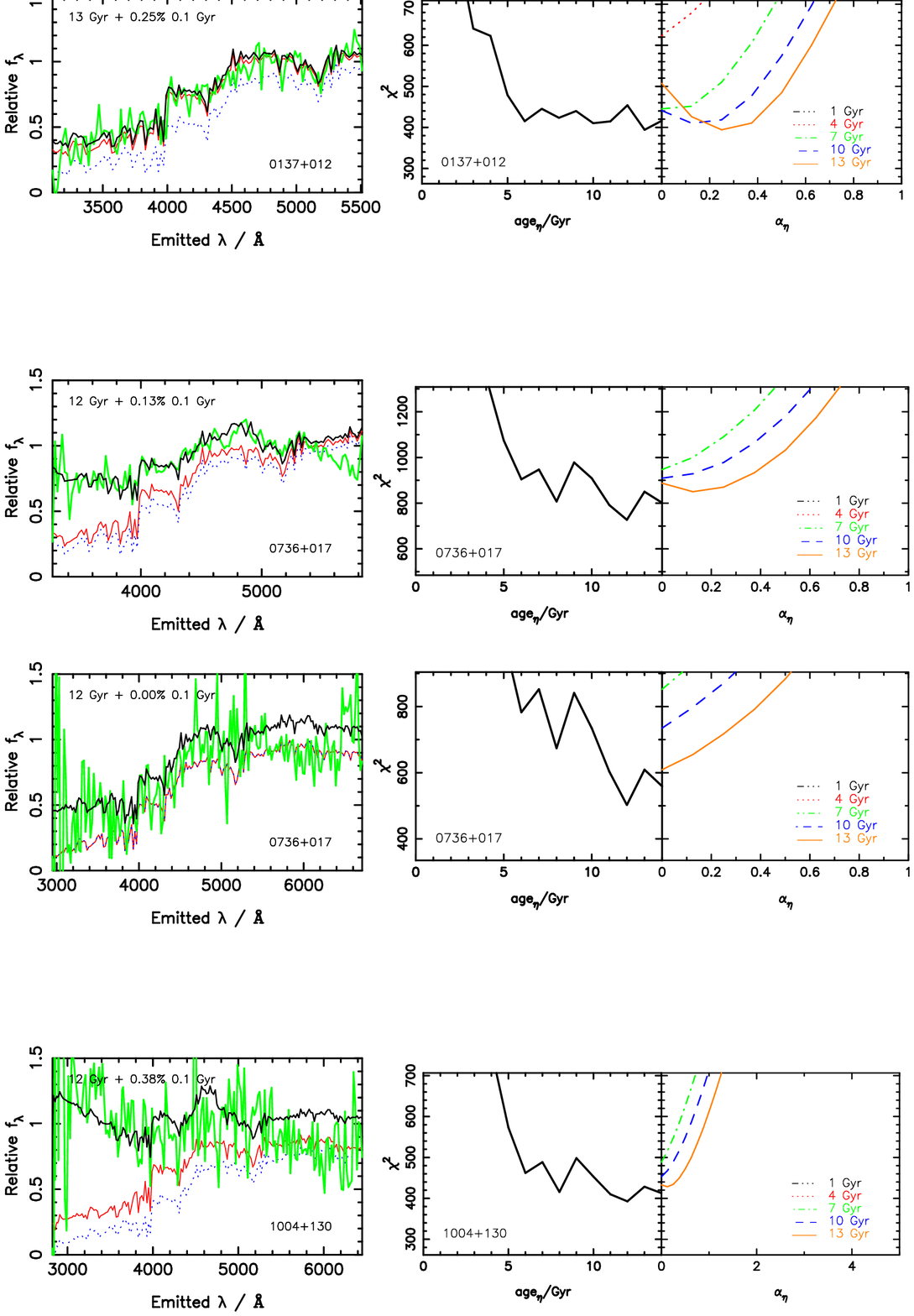,width=15cm,angle=0,clip=}}
	\vspace{-1.5cm}
      	\caption{ Model fits to the off-nuclear rest frame spectra,
including the modelling of a nuclear contribution, for each object, with the corresponding $\chi^{2}$ plots. The rest frame 
host galaxy spectra are in the first column (light grey), with the 
best-fitting two-component model flux plus the nuclear flux contribution (black) 
and the best-fitting two-component model spectra (Jimenez et al. 2000) 
superimposed (mid grey). As in Fig A1, the spectra of the single-aged old 
population (dotted line) is given for comparison. The second column shows 
the \xs evolution with age for the dominant older population and the third 
column shows the best-fit \xs\ as a function of percentage young population, $\alpha$, 
for fixed ages of the dominant component. The subscript $\eta$ denotes 
results obtained by including the nuclear contribution. All models have 
solar metallicity. Where there are two spectra of the same object, the 
spectrum given first is the one observed on the Mayall 4m Telescope, and 
the second is that observed using the William Herschel Telescope.
The data for the following objects have been smoothed using a Hanning 
function: 2135+147 (RLQ), 2141+175 (RLQ), 0244+194 (RQQ), 0923+201 (RQQ), 1549+203 
(RQQ), 2215$-$037 (RQQ), 0230$-$027 (RG) and 0345+337 (RG).}
\end{figure*}
 	
\begin{figure*}
\setcounter{figure}{1}
\centerline{{\LARGE {\em Radio Loud Quasars}} }
\centerline{\epsfig{file=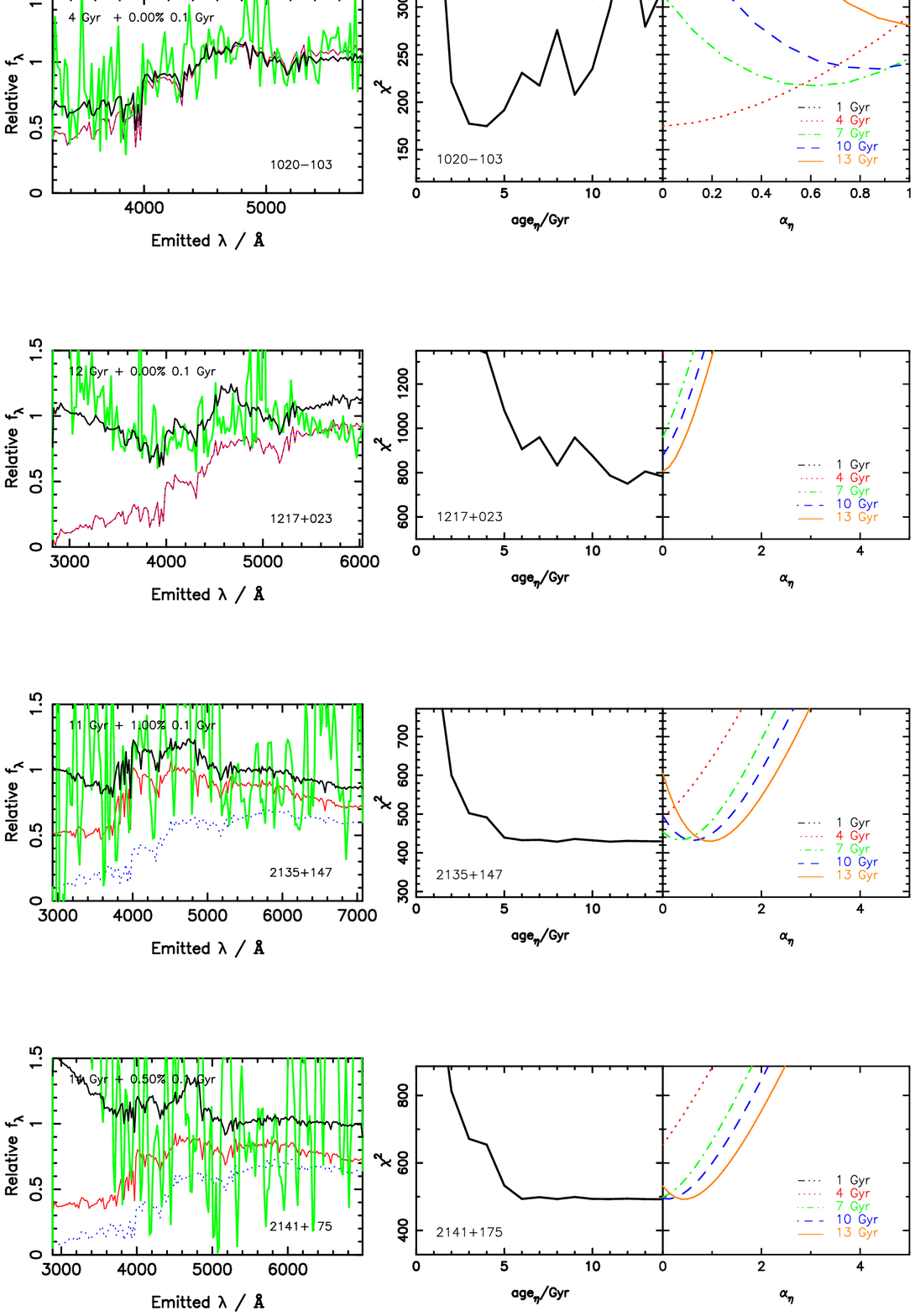,width=16cm,angle=0,clip=}}
	\vspace{-1.5cm}
        \caption{Model fits to the off-nuclear rest frame spectra, including the modelling of a nuclear contribution, continued. }
\end{figure*}

\begin{figure*}
\setcounter{figure}{1}
\centerline{{\LARGE {\em Radio Loud Quasars}} }
\centerline{\epsfig{file=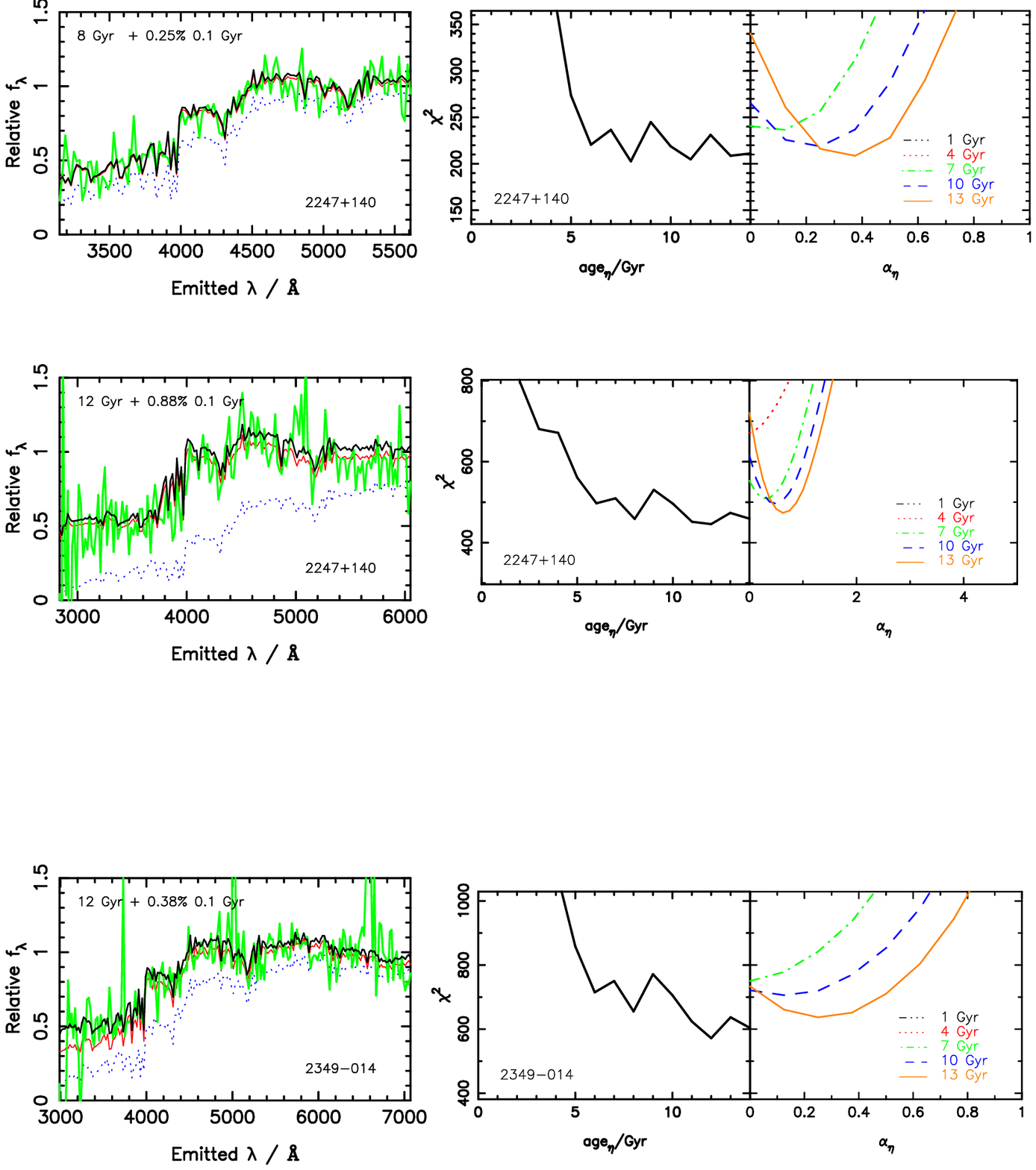,width=16cm,angle=0,clip=}}
	\vspace{-6cm}
        \caption{Model fits to the off-nuclear rest frame spectra, including the modelling of a nuclear contribution, continued. }
\end{figure*}

\begin{figure*}
\setcounter{figure}{1}
\centerline{{\LARGE {\em Radio Quiet Quasars}} }
\centerline{\epsfig{file=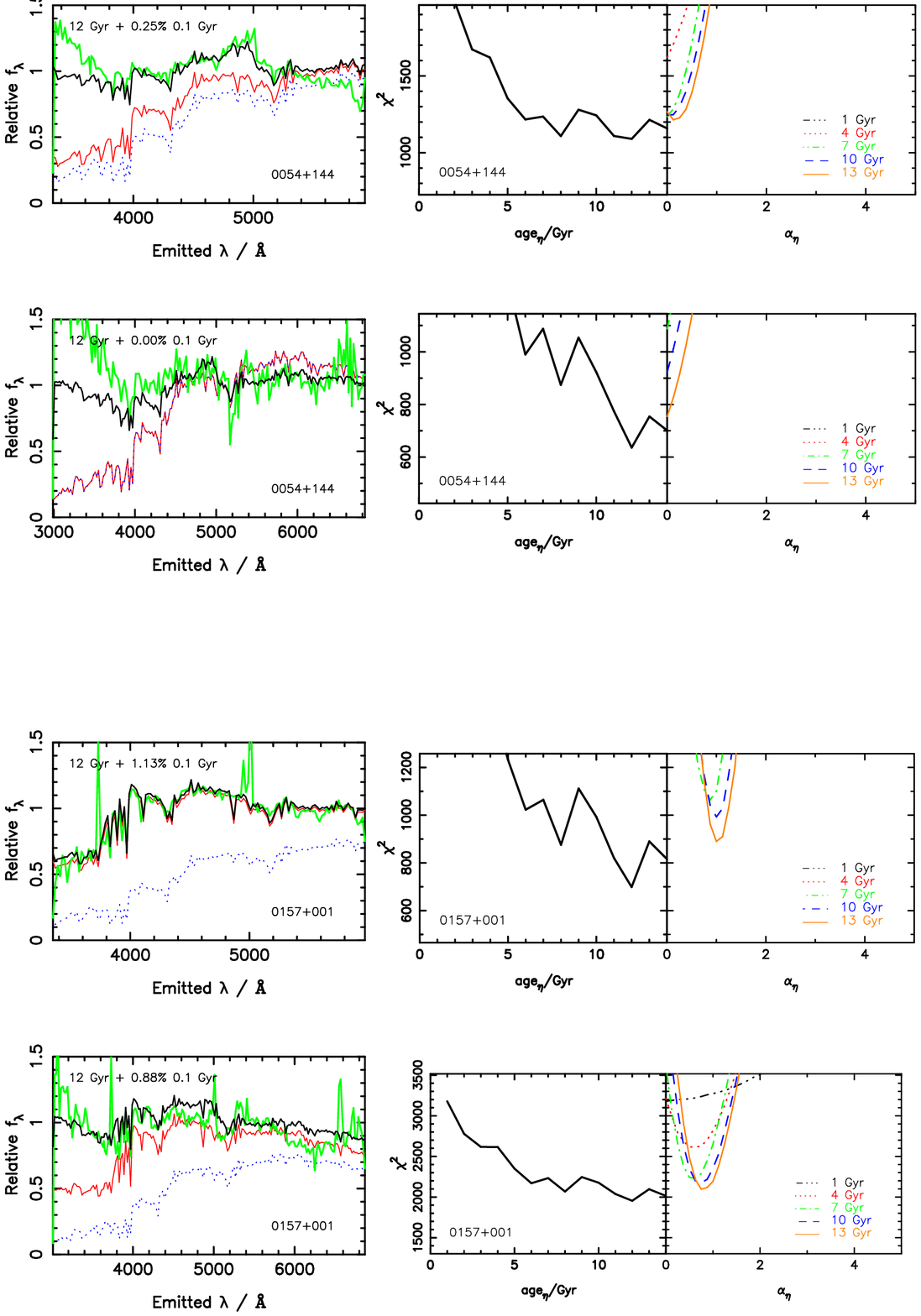,width=16cm,angle=0,clip=}}
	\vspace{-1.5cm}
        \caption{Model fits to the off-nuclear rest frame spectra, including the modelling of a nuclear contribution, continued. }
\end{figure*}

\begin{figure*}
\setcounter{figure}{1}
\centerline{{\LARGE {\em Radio Quiet Quasars}} }
\centerline{\epsfig{file=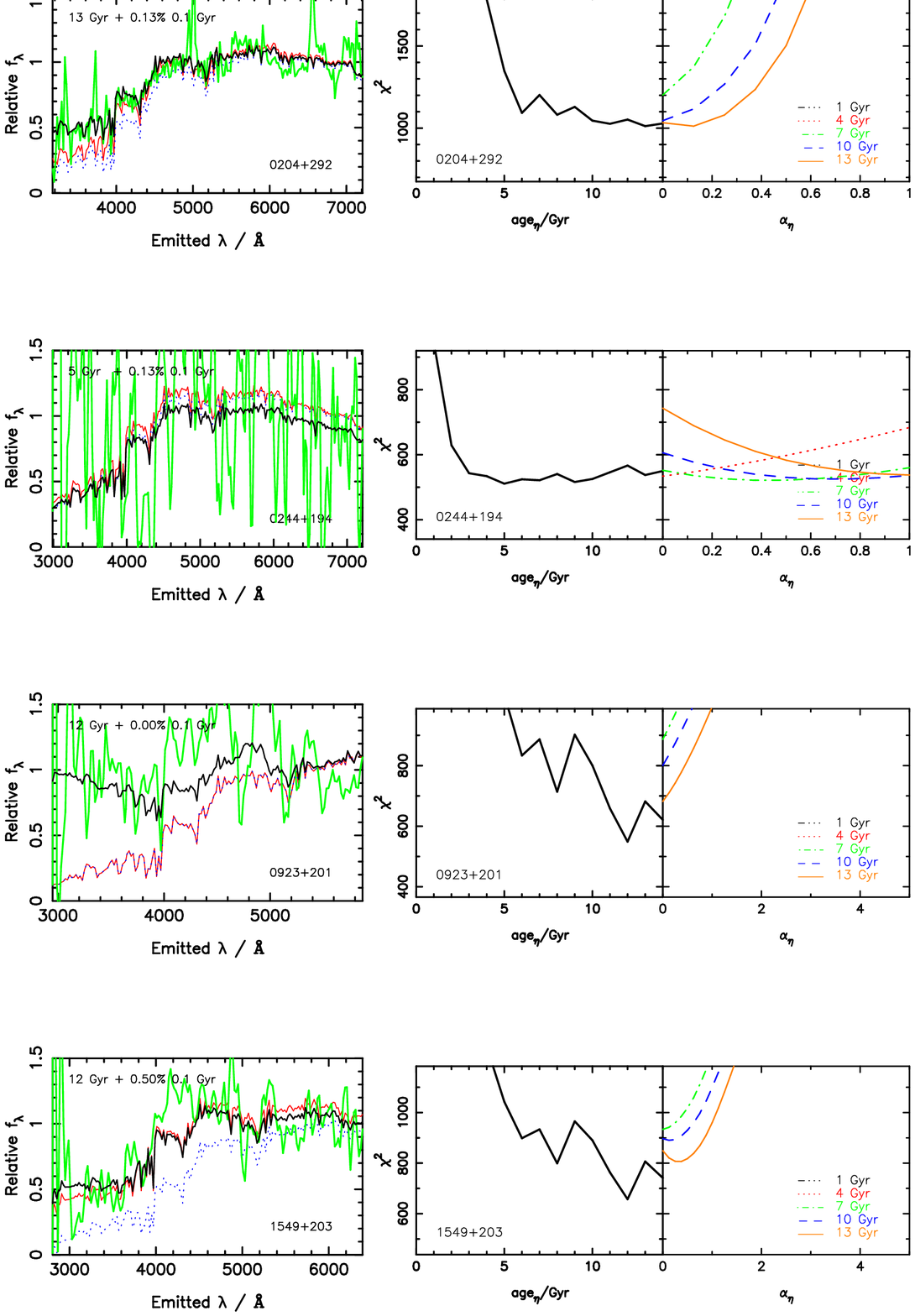,width=16cm,angle=0,clip=}}
	\vspace{-1.5cm}
        \caption{Model fits to the off-nuclear rest frame spectra, including the modelling of a nuclear contribution, continued. }
\end{figure*}

\begin{figure*}
\setcounter{figure}{1}
\centerline{{\LARGE {\em Radio Quiet Quasars}} }
\centerline{\epsfig{file=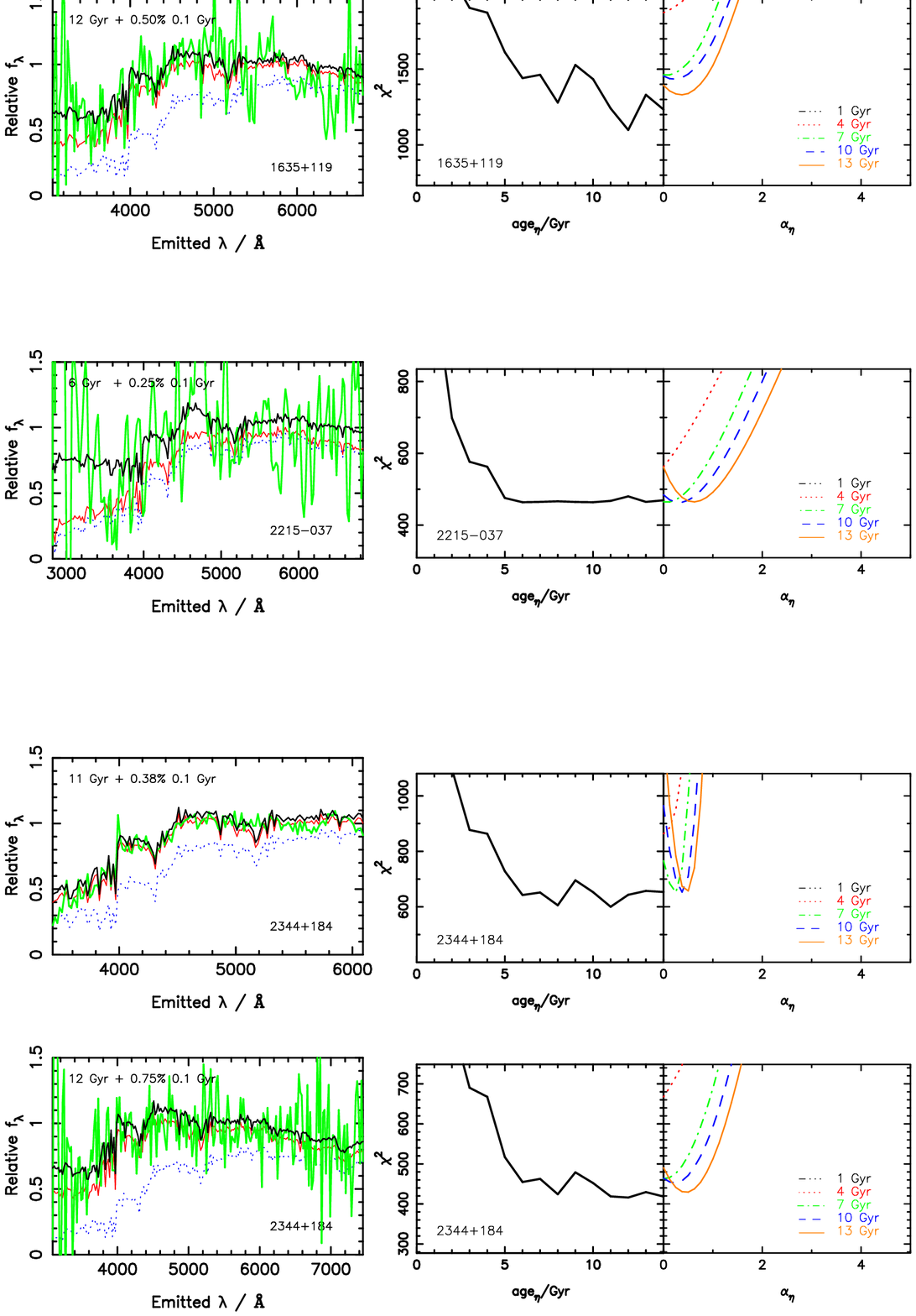,width=16cm,angle=0,clip=}}
	\vspace{-1.5cm}
        \caption{Model fits to the off-nuclear rest frame spectra, including the modelling of a nuclear contribution, continued. }
\end{figure*}

\begin{figure*}
\setcounter{figure}{1}
\centerline{{\LARGE {\em Radio Galaxies}} }
\centerline{\epsfig{file=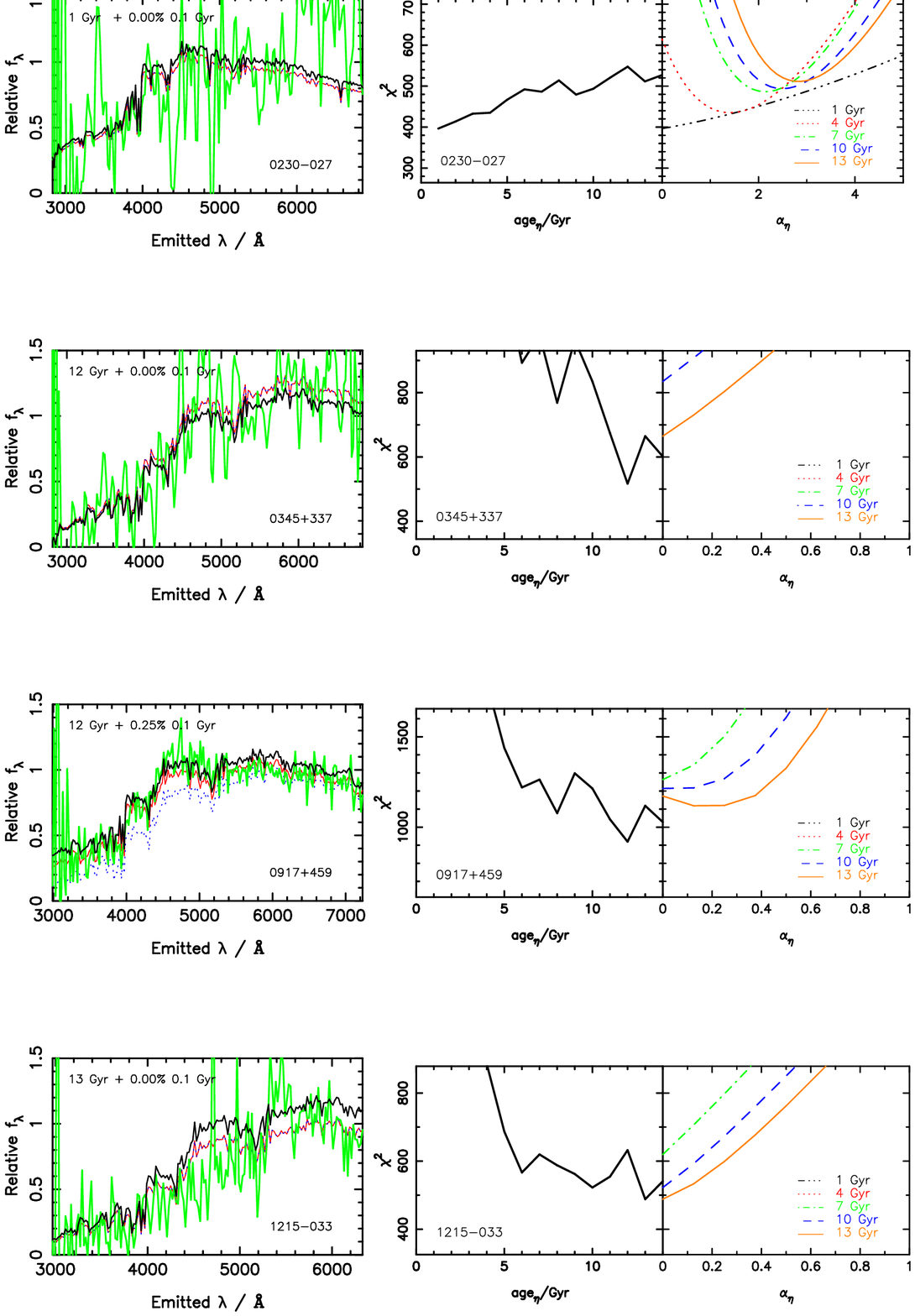,width=16cm,angle=0,clip=}}
	\vspace{-1.5cm}
        \caption{Model fits to the off-nuclear rest frame spectra, including the modelling of a nuclear contribution, continued. }
\end{figure*}

\begin{figure*}
\setcounter{figure}{1}
\centerline{{\LARGE {\em Radio Galaxies}} }
\centerline{\epsfig{file=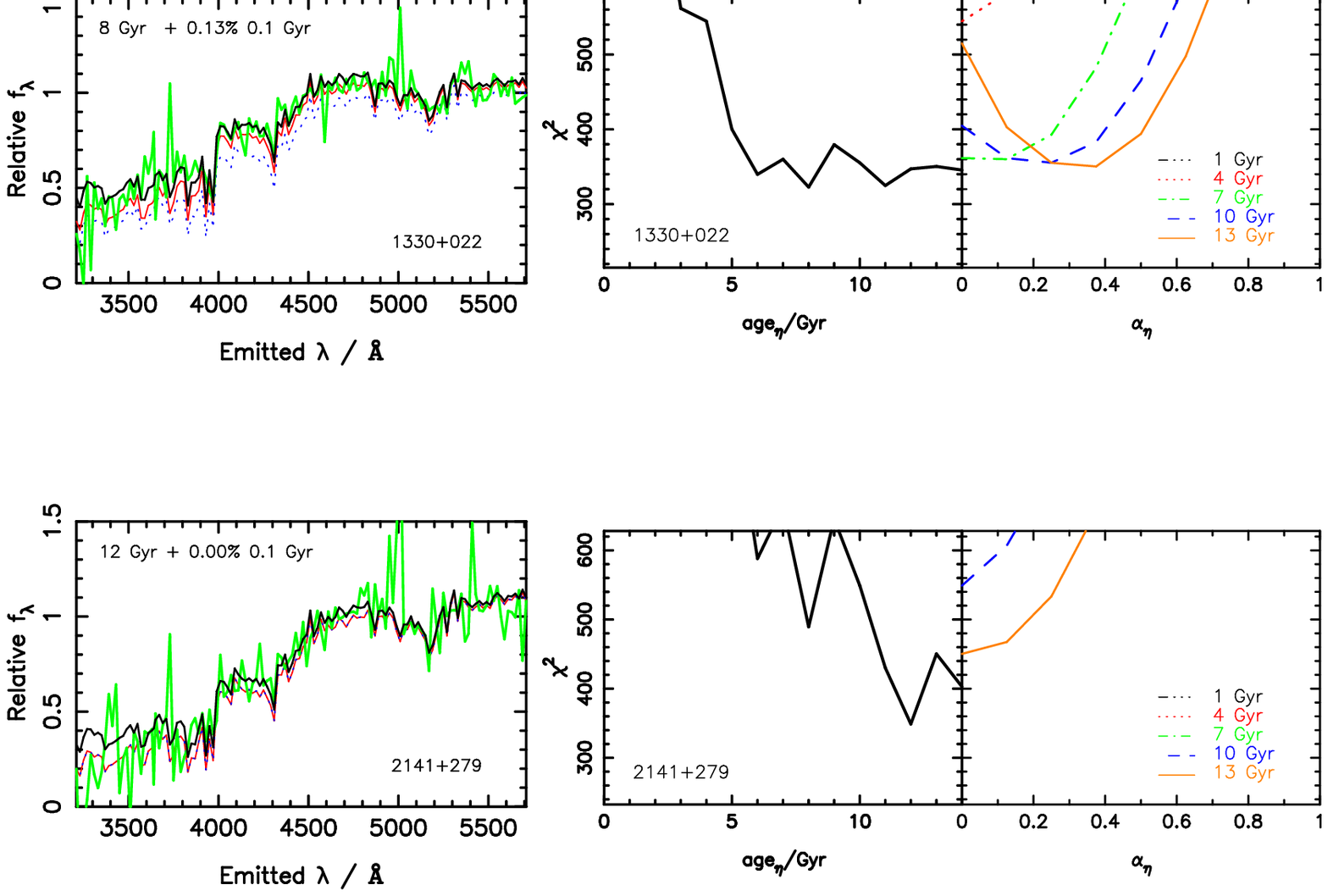,width=16cm,angle=0,clip=}}
	\vspace{-11cm}
        \caption{Model fits to the off-nuclear rest frame spectra, including the modelling of a nuclear contribution, continued. }
\end{figure*}

\end{document}